\documentclass[10pt,twocolumn,letterpaper]{article}

\usepackage{cvpr}              %

\usepackage[dvipsnames]{xcolor}

\usepackage{amssymb}
\usepackage{booktabs}
\usepackage{fancyref}
\usepackage{xcolor}

\usepackage{subcaption}

\usepackage{tikz}
\usepackage{pgfplots}
\usepgfplotslibrary{groupplots}
\pgfplotsset{compat=1.17}

\usepackage{xurl}

\definecolor{cvprblue}{rgb}{0.21,0.49,0.74}

\definecolor{best}{rgb}{1, 0.7, 0.7}
\definecolor{second}{rgb}{1, 0.85, 0.7}
\usepackage[pagebackref,breaklinks,colorlinks,citecolor=cvprblue]{hyperref}
\usepackage{tabularx} %

\def\paperID{*****} %
\def\confName{3DV\xspace}
\def\confYear{2026\xspace}

\title{
  Capture Stage Matting: Challenges, Approaches, and Solutions \\
  for Offline and Real-Time Processing
}

\author{
Hannah Dr{\"o}ge\footnotemark[1],
Janelle Pfeifer\footnotemark[1],
Saskia Rabich,
Reinhard Klein,
Matthias B. Hullin,
Markus Plack \\
University of Bonn \\
Bonn, Germany
}

\begin{document}
\twocolumn[{%
\renewcommand\twocolumn[1][]{#1}%

\maketitle

\vspace{-14pt}
\begin{center}
    \centering
    \captionsetup{type=figure}
    \includegraphics[width=.98\textwidth]{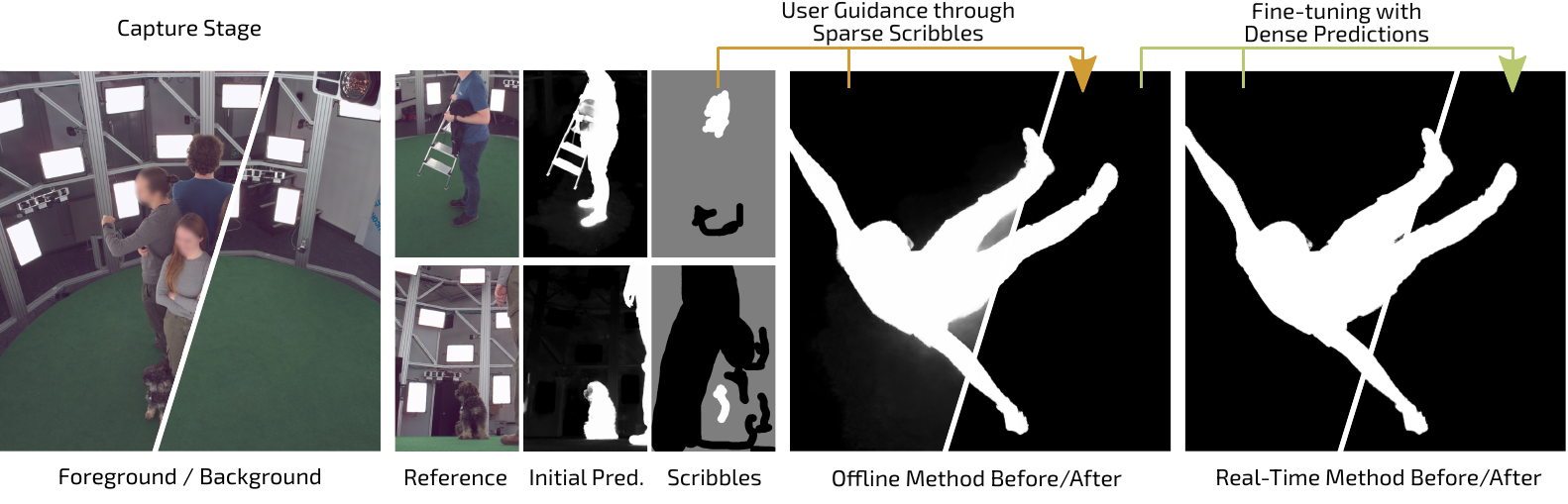}
    \vspace{-6pt}
    \captionof{figure}{We present a pipeline that adapts state-of-the-art matting approaches to the challenges of capture stage setups 
leveraging background information that is readily available in these controlled environments (left). 
We demonstrate how to guide an offline matting method via sparse scribbles (center) and use the improved predictions from the offline method to teach a lightweight real-time matting method (right), enabling better performance across both offline and real-time workflows. \label{fig:teaser}}
\end{center}%
}]
\renewcommand{\thefootnote}{\fnsymbol{footnote}}
\footnotetext[1]{Authors contributed equally.}
\renewcommand{\thefootnote}{\arabic{footnote}}

\begin{abstract}
\vspace{-0.4cm}
Capture stages are high-end sources of state-of-the-art recordings
for downstream applications in movies, games, and other media.
One crucial step in almost all pipelines is matting, i.e., separating captured performances from the background.
While common matting algorithms deliver remarkable performance
in other applications like teleconferencing and mobile entertainment,
we found that they struggle significantly with the peculiarities
of capture stage content.
The goal of our work is to share insights into those challenges 
as a curated list of these characteristics along with a constructive discussion for proactive intervention 
and present a guideline to practitioners for an improved workflow 
to mitigate unresolved challenges.
To this end, we also
demonstrate an efficient 
pipeline to adapt
state-of-the-art approaches to such custom setups without
the need for extensive annotations,
both offline and real-time.
For an objective evaluation, we introduce a validation methodology
using a state-of-the-art diffusion model
to demonstrate the benefits of our approach.

\end{abstract}

\section{Introduction}
\label{sec:intro}

Matting is a fundamental technique in the field of computer vision, playing a key role in tasks requiring precise foreground-background separation. 
Mathematically, matting can be expressed by a linear combination of fore- and background images $F$ and $B$, enabling an alpha blending between foreground and background,
\begin{equation} \label{eq:linear_combination}
    I_i = \alpha_i F_i + (1-\alpha_i) B_i,\; \alpha_i \in [0,1],
\end{equation}
where $\alpha_i$ represents the desired pixel-wise transparency value coefficient.
It is typically performed in the early stages of computational workflows, making subsequent calculations highly sensitive to errors in foreground separation.
For instance, matting serves as a key preprocessing step in reconstruction methods using the visual hull \cite{plack2024vhs} as well as hair reconstruction techniques, where precise separation of the hair and head from the background is required \cite{sklyarova2023neural}.

Over the last decade, deep learning methods have driven the development of state-of-the-art matting techniques capable of fine-grained transparency estimation, enabling the application of matting techniques in a wide range of fields, including augmented reality and film production. %
An important application domain lies within capture stages that provide multi-camera setups for recording high-quality data in controlled environments.
Robustness of matting is essential for reconstructing high-fidelity content, as large volumes of multi-camera data make per-frame supervision infeasible.
However, achieving such matting in these setups remains a significant challenge, as state-of-the-art methods \cite{hu2025diffusion, yao2024vitmatte, tang2019learning} typically rely on predefined trimaps or large-scale labeled datasets, both impractical for capture stages:
Trimaps demand highly accurate per-frame generation, and existing datasets generalize poorly to these environments, often leading to errors.

Fortunately, capture stages characteristically feature a static background and controlled lighting, which enable the capture of the empty stage as a reference for matting,
providing additional information to support the process.
However, this approach introduces new challenges, as reflective materials (e.g., structures within the stage) and soft shadows cast by the subject can significantly alter the appearance of the stage in each shot, complicating the matting workflow, especially given the persistent lack of high-quality training data.

In this work, we address these challenges by discussing the potential failure cases in capture stage setups, providing a guideline for improved matting in these environments, and improving deep matting techniques through a pipeline specifically designed for such capture stage setups. In particular, we place a special focus on real-time matting, which is highly relevant for streaming and telepresence applications.
In this context, we discuss a two-phase pipeline based on a student-teacher concept. 
We modified 
an offline matting model to apply background matting %
 and adapted it to process capture stage data using coarse user-drawn annotations in the first phase.
 This adapted model serves as a teacher, whose refined outputs guide a real-time matting model
 to generate high-quality mattes for capture stage data by learning from the refined outputs of the teacher{. We} {evaluate both models using a hand-labeled trimap-based diffusion model on a hold-out validation set.}
 In summary, our contributions are:
\begin{itemize}
  \item A qualitative analysis of challenges and potential failure cases of (deep) matting methods in capture stage environments (\cref{sec:problem_setup})
    \item We present a lightweight, yet efficient pipeline to improve learned matting approaches for known but potentially wrong background information, both offline and in real-time (\cref{sec:guide})
  \item We validate our approach against a precise, trimap-based matting method and present the effects of accurate matting in a NeRF-based downstream application, showing improvements in rendering quality (\cref{sec:evaluation})
\end{itemize}

\section{Related Work}
\label{sec:rel_work}

With the rise of deep learning, Xu \etal \cite{xu2017deep} introduced a deep image matting model and
released a dataset enabling training of models for alpha mask prediction.
Since then, numerous works {applied} learning methods for deep image matting, many of which rely on pre-defined trimaps \cite{yao2024vitmatte, hu2025diffusion}, that often present a bottleneck due to the manual labeling required for their creation. 
Zhang \etal \cite{zhang2024learing} addressed this challenge by learning trimaps from sparse user input, while others \cite{lyu2024privileged} developed a student-teacher model, letting a trimap-free student model learn from a trimap-guided teacher. Hachmann \etal \cite{hachmann2023color} replaced the trimap completely by predicting the matting based on  successive images of a scene with changing background colors. 
To facilitate the matting task, some works {built} upon segmentation models \cite{li2024matting, yao2024matte}, or used them for additional annotation \cite{kim2024towards}, 
while another line of research {used} attention mechanisms and transformer models for matting \cite{yao2024vitmatte, cai2023transmatting}, e.g. by using cross-attention to capture
object contours
\cite{wang2024eformer}. As diffusion models gained popularity, they have been increasingly utilized in the field of deep image matting~\cite{wang2024matting, guo2024context}. 

Research has focused on improving matting by optimizing training data via better foreground-background composition \cite{ye2023infusing, li2022bridging}, while self-supervised pretraining on random trimaps and alpha masks has been introduced for better model initialization \cite{li2024disentangled}. Other work reduces annotation effort by proposing a novel loss that trains exclusively on trimaps \cite{liu2024training}. A dataset generation method has also been developed to produce images tailored for chroma keying \cite{burgert2024magick}. In the context of domain shift, image harmony has been used to narrow the gap between real and synthetic data \cite{zhao2024boosting}, and a new foreground–background composition strategy has been proposed to reduce the discrepancy between those, without requiring prior information \cite{li2022bridging}.

Several works proposed real-time approaches for image and video matting to support certain downstream applications like live-streaming. In this context, Lin \etal~\cite{lin2021real} {presented} a real-time matting model that uses additional background information for image matting, while other methods \cite{ke2022modnet, qin2023bimatting, lin2022robust} solely rely on the image containing the foreground object.
Among these methods, Ke \etal~\cite{ke2022modnet} proposed \textit{ModNet}, a real-time matting model specifically designed for portrait matting, addressing flickering issues caused by the lack of temporal coherence in video matting. 
BiMatting \cite{qin2023bimatting} applies mask binarization to meet  real-time requirements, while a separate recurrent-network solution \cite{lin2022robust} enhances robustness for continuous video matting.
In this context, Huynh \etal~\cite{huynh2024maggiemaskedguidedgradual} proposed a mask-guided method for human instance matting, while Lin \etal~\cite{lin2023adaptive} developed a transformer-based approach for matting humans in unstructured video environments.
For a comprehensive overview of deep image matting methods, please refer to the survey by Li \etal~\cite{li2023deep}.

\section{Problem Setup and Challenges}
\label{sec:problem_setup}

Capture stages are designed to maximize the quality of recordings
but can still pose significant challenges to learned matting approaches in particular.
While the readily available background information that can be trivially acquired by capturing an empty stage
is a major benefit,
we have identified several intricacies of those environments
that complicate a successful foreground separation.
Considering the composite image $I_i$ in \cref{eq:linear_combination}, described as a linear combination of the foreground content \(F_i\), and the background \(B_i\), practical setups inherently introduce complexities that deviate from this idealized model. 
Noise, environmental interactions, and sensor limitations in real-world scenarios result in alterations that demand a more complex model,
\begin{equation} \label{eq:lineq_art}
    I_i = \alpha_i F_i + (1-\alpha_i)\,\bigl(B_i + \delta_B(F)\bigr) + \varepsilon_i,\quad \alpha_i \in [0,1],
\end{equation}
whereby $\delta_B(F)$ represents alterations to $I_i$ caused by the influence of the foreground subject's shadows, reflections, or color bleed, while $\varepsilon$ represents sensor noise and other unmodeled artifacts.

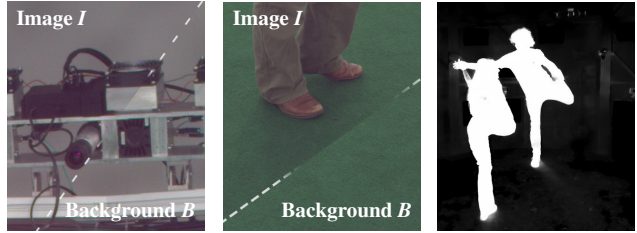
\begin{figure}
    \centering
    \begin{minipage}[b]{0.15\textwidth}
        \begin{tikzpicture}
            \node[inner sep=0pt] (img) {\includegraphics[width=\textwidth]{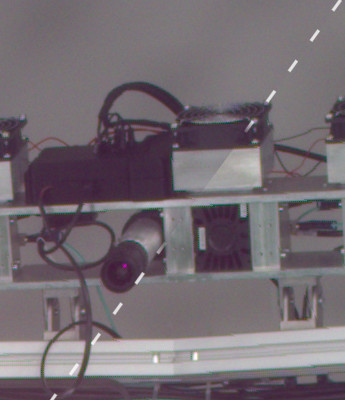}};
            \node[anchor=north west, text=white, font=\bfseries\footnotesize] at (img.north west) {Image \textit{I}};
            \node[anchor=south east, text=white, font=\bfseries\footnotesize] at (img.south east) {Background \textit{B}};
        \end{tikzpicture}
    \end{minipage}%
    \hspace{0.1em} %
    \hfill
    \begin{minipage}[b]{0.15\textwidth}
        \begin{tikzpicture}
            \node[inner sep=0pt] (img) {\includegraphics[width=\textwidth]{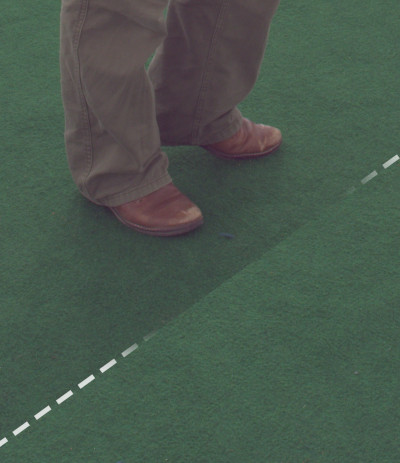}};
            \node[anchor=north west, text=white, font=\bfseries\footnotesize] at (img.north west) {Image \textit{I}};
            \node[anchor=south east, text=white, font=\bfseries\footnotesize] at (img.south east) {Background \textit{B}};
        \end{tikzpicture}
    \end{minipage}
    \hfill
    \begin{minipage}[b]{0.15\textwidth}
        \includegraphics[width=\textwidth]{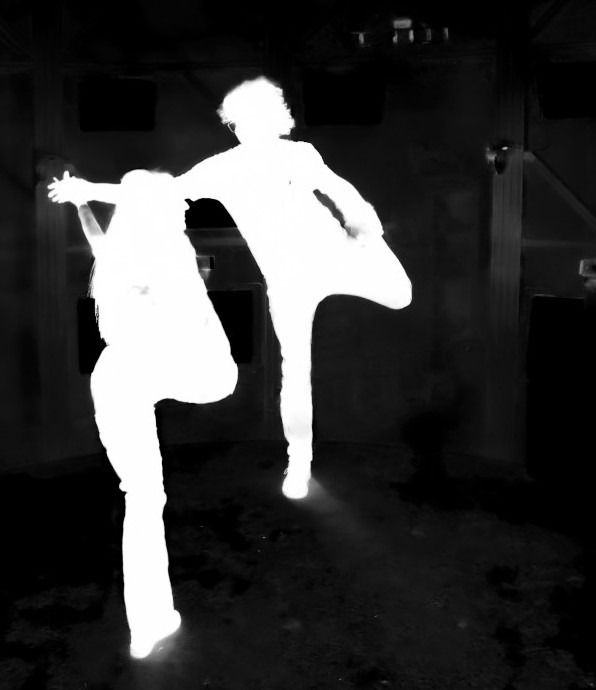}
    \end{minipage}
    
    \caption{Challenging conditions, like reflective  materials of the capture stage setup (left), as well as cast shadows (center), can lead to alterations of the background image and significantly impact matting (right).}
    \label{fig:challenges}
\end{figure}

Even in highly controlled studio conditions, minor deviations between the “ideal” background $B$ and the actual recordings of the stage with a foreground subject present are inevitable.
These \textbf{background alterations} in $I$ are reinforced in particular by reflective materials such as the housings of cameras, lenses or metal frames, which reflect light and the subject. 
Shadows cast by the subject further alter the appearance of the background, whereby soft shadows occur especially at the feet of a subject, as shown in \cref{fig:challenges}. Reflections and shadows make the pre-captured background information unreliable, leading to ambiguity in distinguishing the fore- and background. 
These deviations in the background can substantially affect the accuracy of alpha estimation, as the matting algorithm may fail to attribute background changes correctly to either the foreground layer \(F\) or the background layer \(B+ \delta_B(F)\).

Like all camera setups, capture stages are subject to \textbf{noise} $\varepsilon$, which effectively alters the appearance of the background in each recording. 
Noise degrades the edges of objects, making pixel-accurate distinctions difficult, 
misleading a matting network into making incorrect predictions.
While longer exposure times reduce noise, they also increase undesirable motion blur, particularly when capturing a moving person. In a capture stage setup, this can either require uncomfortably bright lights or high gain levels which further increase the noise and create more pronounced differences in the background.

\begin{figure}
  \centering
  \includegraphics[width=0.23\textwidth]{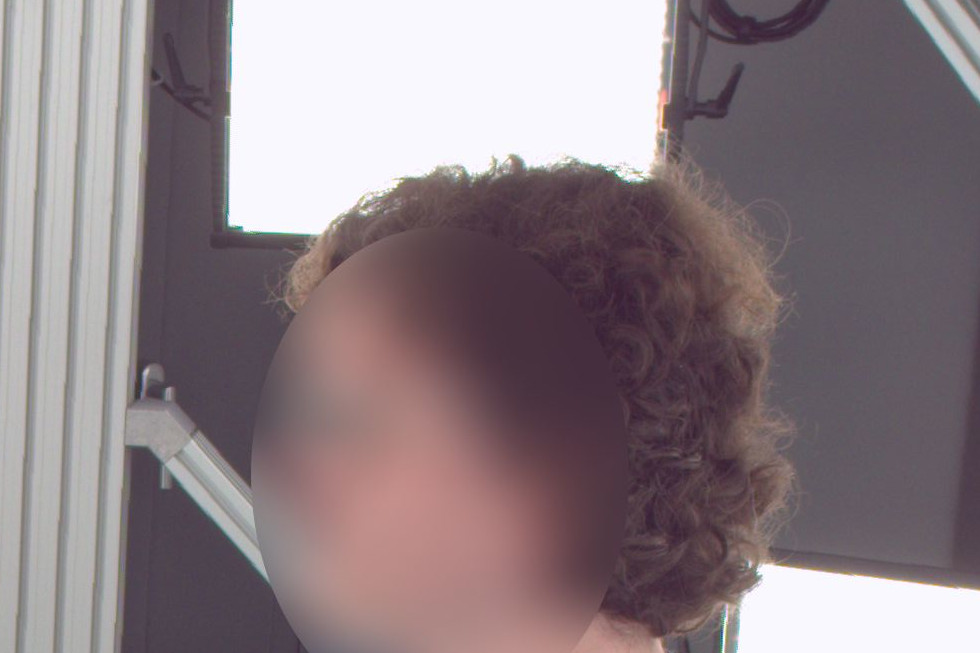}
  \includegraphics[width=0.23\textwidth]{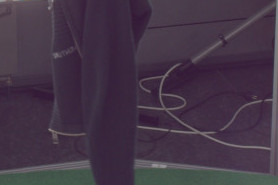}
  \caption{
    Ambiguities in foreground and background information can be caused by lighting in the background (left) and dark objects against similarly dark backgrounds (right).
\vspace{-0.3cm}
  }
  \label{fig:ambiguities}
\end{figure}

\begin{figure*}
    \centering
    \includegraphics[width=0.99\textwidth]{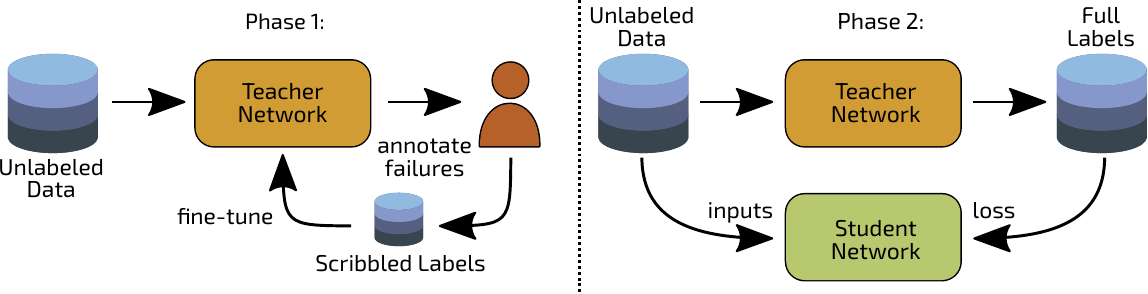}
    \caption{Illustration of the student-teacher model refinement process. The teacher network is fine-tuned on specific failure cases provided by the user to improve its predictions, even on unseen datasets (left). The real-time student model is subsequently fine-tuned based on corrected predictions of the improved teacher model, ensuring better generalization and performance (right). }
    \label{fig:teacher-student-refinement}
\end{figure*}
Beyond the challenges of exposure settings, the interplay between lighting conditions and the captured subject introduces \textbf{ambiguities}, which can cause fine foreground structures, such as hair, to disappear. 
Additionally, areas with minimal contrast, such as dark objects against similarly dark backgrounds, further complicate the task of accurate distinction, as illustrated in 
\cref{fig:ambiguities}. %

Overcoming such challenges is not only a technical challenge but also highlights the crucial role of high-quality labeled \textbf{training data} in improving matting algorithms.
Creating such data for matting requires the creation of alpha masks, which is labor-intensive, time-consuming, and error-prone. It typically involves either the use of specialized equipment or manual labeling, which is impractical for large datasets. The lack of high-quality labeled data of capture stage setups limits the ability to train and evaluate matting on such data. 

Also, while matting models are generally trained on synthetic or carefully prepared datasets of real-world occurrences, this data {does not cover the conditions observed in capture stage environments}. The latter differ in lighting condition, background, and texture, leading to a significant \textbf{domain gap}, which results in decreased performance and reliability when matting models are applied to capture stage data.

The challenges introduced by the domain gap are further complicated in the context of  
\textbf{real-time matting methods}, which prioritize speed and efficiency, often at the expense of performance. Due to their limited capacity, these methods may fail to capture details of the foreground subject, especially in complex scenarios, and can be expected to exhibit lower generalization performance which limits their capabilities to incorporate sparse additional inputs, such as scribbles, for post-training refinements.

\section{A Guide to Better Matting}
\label{sec:guide}

{Building on the challenges outlined in the previous chapter, we present approaches based on our own experiences as guidelines
for improved matting in capture stage environments, including an optimized hardware setup, and a student-teacher approach  that incorporates targeted refinement strategies, as illustrated in \cref{fig:teacher-student-refinement}.} %

\subsection{Start with a Better Setup}

Perhaps the most crucial point of optimization lies long before the data processing
during the planning of the capture stage itself.
For the construction of the stage, we have found that glossy materials should be avoided wherever possible. For example, carpet instead of hardwood flooring creates a diffuse surface and has the added benefit of improved safety against slipping. Diffuse materials should also be preferred for the housing of cameras and scaffolding holding the cameras and lights.
It is also beneficial to block the view from the capture stage into the room. This could be done with fabric or frosted plexiglass. The goal is a mostly uniform background color. When choosing lights, it is important to consider that small details like hair will disappear when placed in front of bright lights.
Spreading the lighting out over a larger area or customizing the light positions according to the capture subject may lead to improvements.
Some issues, however, like soft shadows, cannot be solved through the setup and need to be addressed through software.

\subsection{Error Aligned Fine-tuning via Sparse Labels}
\label{sec:scribble_labels}

Large state-of-the-art networks generally deliver high-quality results for \textit{most} images, failing significantly in only a few cases. Given a large enough training dataset, we assume that current networks could easily learn to overcome those problems. Therefore,  we aim to extend the existing training dataset with images of the capture stage setup where matting fails.
Starting with a pretrained \textit{teacher} network, as shown in the first phase in  \cref{fig:teacher-student-refinement}, the model produces initial predictions on a given set of images taken within the capture stage environment. Based on these predictions, failure cases are  annotated by users with scribbles (compare \cref{fig:teaser}).
This process creates a  small dataset, that highlights the problematic areas for further refinement.
The matting network is fine-tuned on a combination of the original dataset and the newly annotated capture-stage samples. For the latter, only the annotated regions are included in the loss during training.
This approach allows the network to address the challenges of the capture stage data by using the annotated regions for targeted improvements.
Meanwhile, the original dataset ensures that the network maintains a strong baseline performance across a diverse range of scenarios.
This process can be repeated iteratively to achieve increasingly reliable results as needed.

\subsection{Student-Teacher Learning for Fast Matting}

A \textit{student-teacher} framework is ideal for addressing the performance gap between offline and real-time processing. 
As shown in the second phase in \cref{fig:teacher-student-refinement}, the high-capacity network that has been fine-tuned on the coarse annotations (\cref{sec:scribble_labels}),  serves as the \textit{teacher} model, generating high-quality alpha masks on a larger dataset of unlabeled capture-stage data. 
These teacher-generated outputs are then used as training data to fine-tune a lightweight, real-time \textit{student} network.
This allows the student model to replicate the structural details that the fine-tuned teacher model has captured while preserving the computational efficiency required for real-time applications. 
We observed that directly training the lightweight student model on scribble-labeled data reduces its ability to capture fine details, as shown in \cref{sec:student_train_direct},
which makes intuitive sense, since scribbles usually avoid boundaries, depriving the model of the fine-grained edge information necessary for accurate boundary reconstruction,
highlighting the need for a student-teacher approach.

\subsection{Quality Control}

Although the previous approach already improves robustness, it remains somewhat error-prone due to the absence of quantitative quality control.
To address this limitation, we propose to introduce a validation
mechanism
using a \textit{supervisor} model
to evaluate both the offline and the real-time methods.
The requirements of this model are slightly different, as quality becomes the highest priority, while computational cost and effort required by the user are less important. 
Given those requirements it makes sense to use a trimap-based matting approach, effectively limiting the range of errors within a small window around the edges, which can typically be near-perfectly retrieved by those methods, at the cost of much longer manual labeling and processing times.

\section{Experimental Evaluation and Analysis}
\label{sec:evaluation}

We assess the effectiveness of our approach through both quantitative metrics and qualitative comparisons 
and provide an overview of the used hardware components as well as model specifications and training details.

\subsection{Experimental Setup}

We describe the experimental setup, including hardware configuration, model adaptations, and training procedures.

\paragraph{Hardware Configuration}
Our capture stage features $40$ RGB cameras ($24.5$ megapixels each, downsampled $2\times$ for most applications) and $46$  video lights affixed to cylindrical scaffolding with a diameter of $5.25$ meters. The cameras are capable of capturing $75$ frames per second with a sensor depth of $12$ bits. Subjects in the stage are lit by all video lights and captured by all cameras simultaneously. 
The brightness of the lights in the stage can be controlled, and the camera positions are static, therefore the backgrounds of the cameras do not change and can be pre-captured. 

\begin{figure}[t]
  \centering
  \includegraphics[width=0.23\textwidth]{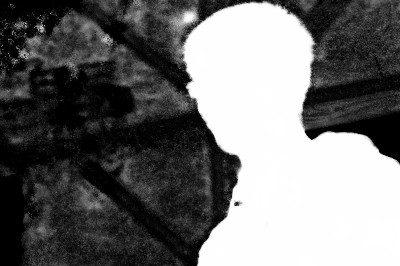}
  \includegraphics[width=0.23\textwidth]{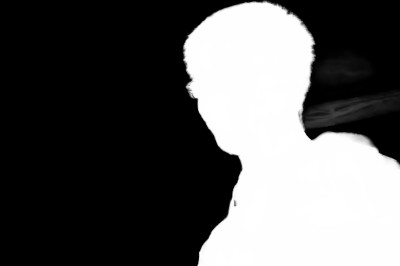}
  \caption{
    Influence of noise in the input images on matting. The left image shows a matting result without introducing noise during training, whereas the right image shows the result with noise added to the augmentation phase during training. 
  }
  \label{fig:noise}
\end{figure}

\paragraph{Implementation Details}
For evaluation, we modified \mbox{\textit{ViTMatte-S}} \cite{yao2024vitmatte}, referred to as \textit{Bg-ViTMatte}, which serves as the offline teacher model in our experiments. 
This included the substitution of the input trimap images with background images, introducing Gaussian noise with a standard deviation of up to $0.1$ during training to adapt the model to our particular problem setup, and train{ing} it on the Adobe Deep Image Matting dataset~\cite{xu2017deep}.
The positive impact of the noise augmentation is demonstrated in \cref{fig:noise}.

\begin{figure}
    \centering
    \includegraphics[width=0.45\textwidth]{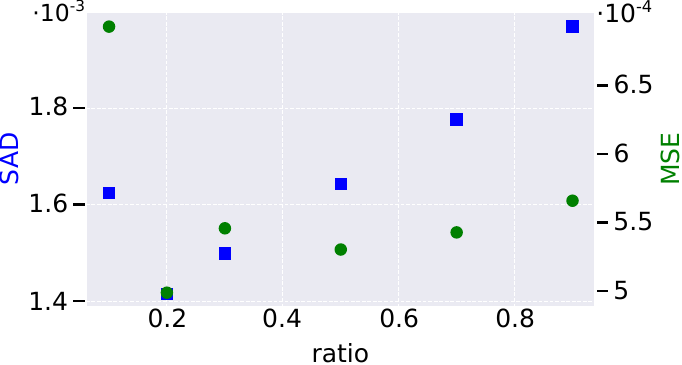}
\caption{The ratio of sparsely annotated data during training the teacher model significantly affects performance. Training exclusively on scribbled data is not included in the plotted range, resulting in an MSE of $78\cdot 10^{-4}$ and an SAD of $11\cdot 10^{-3}$.\label{fig:ratio}}
\end{figure}

In the first phase of our student-teacher refinement process (see \cref{fig:teacher-student-refinement}) we optimized the model on a hybrid dataset -- combining the base dataset with manually annotated $41$ failure cases from capture stage content -- using the loss 
 of~\cite{yao2024vitmatte} for the former and {proposing} a selective loss to focus on relevant regions for the latter,
\begin{equation}
    \mathcal{L}(M,Y) = \; \sum_{i \in \mathcal{S}} |M_i - Y_i|, \text{ with }M = \mathcal{N}_\theta(I, B),
\end{equation}
where $\mathcal{N}_\theta$ represents the matting model, $Y$ the user drawn annotation for image $I$, and $\mathcal{S} = \{i \;|\; Y_i \text{ is annotated}\}$ defines the scribble locations.
We emphasize that the ratio of the combined dataset significantly influences the matting performance, whereas fine-tuning solely on failure annotations leads to poor  predictions, as we demonstrate in \cref{fig:ratio}. Best performance is achieved with $80\%$ of each batch drawn from the base dataset, while the remaining $20\%$ are sampled from the manually annotated failure cases from capture stage content.

We trained Bg-ViTMatte for $2000$ iterations with a batch size of $16$, using an initial learning rate of $5\cdot 10^{-5}$, scheduled to $2.5\cdot 10^{-5}$ after $600$ iterations.
The refined teacher model generates
pseudo-labels
for a new capture stage dataset comprising $20$ scenes with $29$ different camera-views each ($580$ images in total), that are used to fine-tune the student model, \textit{BackgroundMattingV2} (BGMV2) \cite{lin2021real} with a MobileNetV2 backbone, for real-time performance. BGMV2 employs a two-stage architecture: a base model that generates a coarse mask from a fore- and background image, and a second model that improves fine details along the mask's borders. We fine-tuned the base model for $5$ epochs, and the combined model for an additional $10$ epochs using the capture stage dataset and the generated
pseudo-labels.
We have kept all other hyperparameters and losses unchanged, and refer to \cite{lin2021real} for more details.
Validation is conducted using DiffMatte \cite{hu2025diffusion} to create a high-detailed benchmark of 14 high-resolution masks based 
on hand-drawn trimaps from various scenes and camera positions in the capture stage.


\newcommand{\mywidth}{0.15\textwidth}

\begin{figure*}[t]
    \centering
    \begin{subfigure}[t]{\mywidth}
        \centering 
        \includegraphics[angle=270,width=\textwidth]{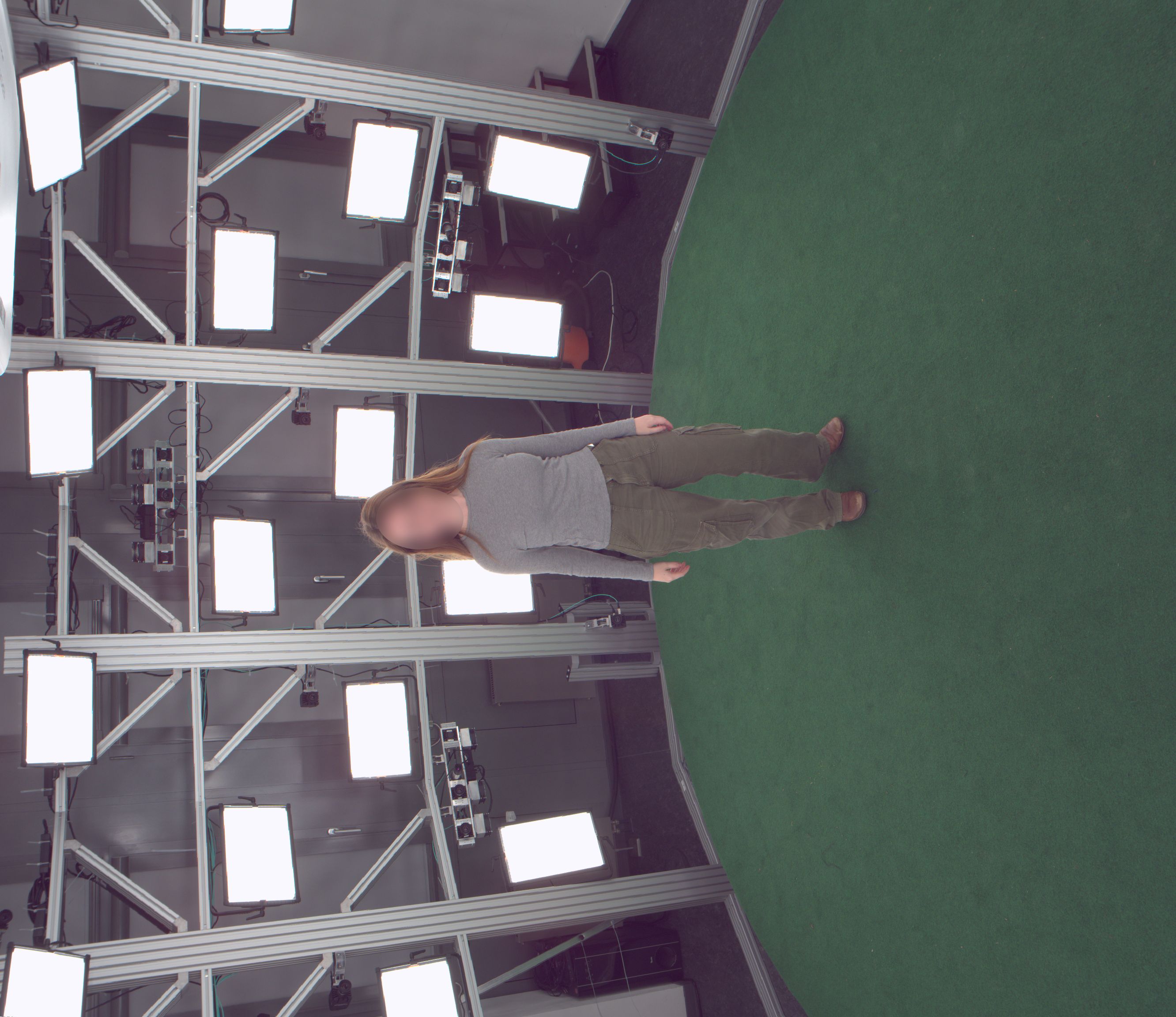}
        \includegraphics[angle=270,width=\textwidth]{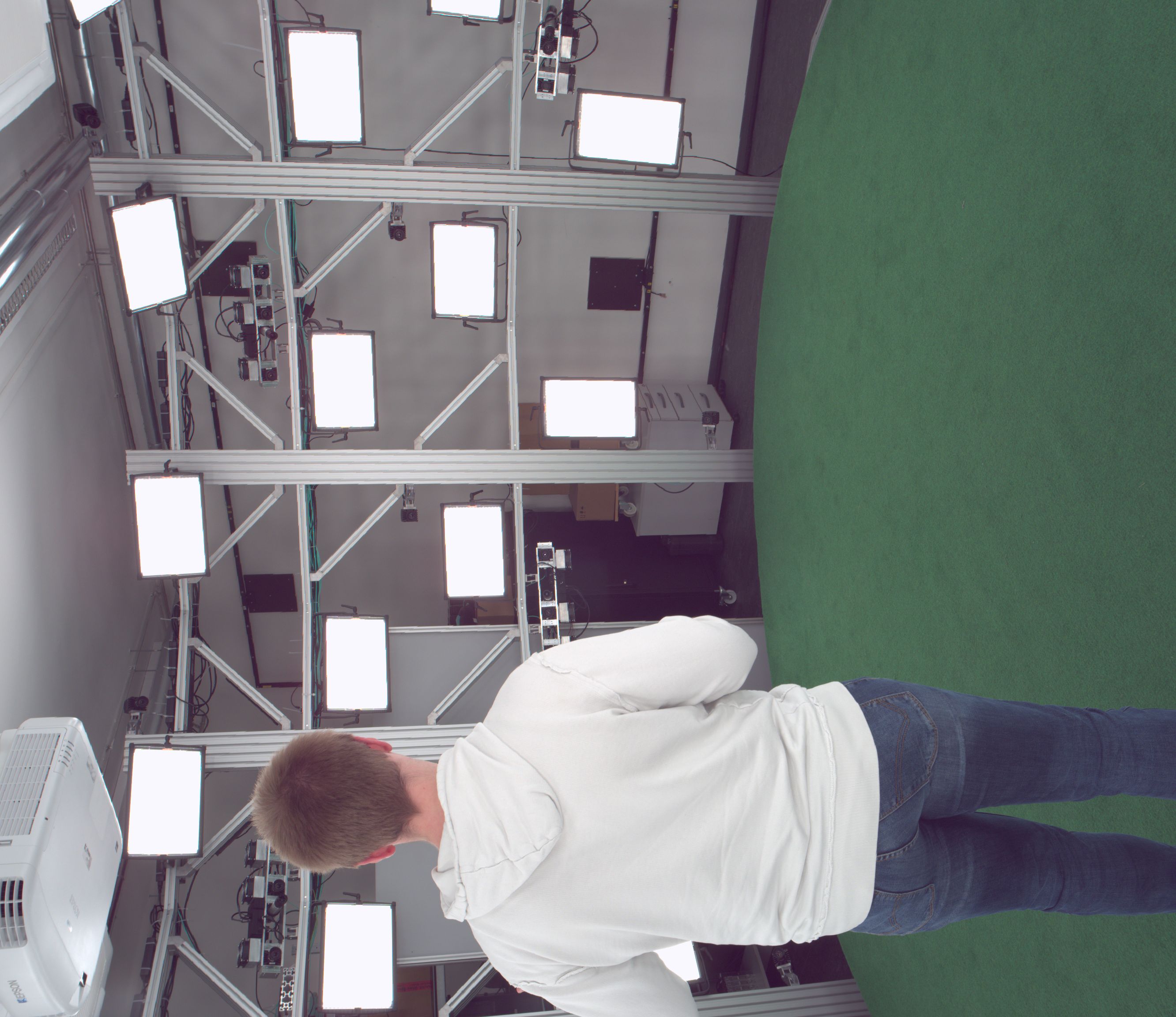}
        \includegraphics[angle=90,width=\textwidth]{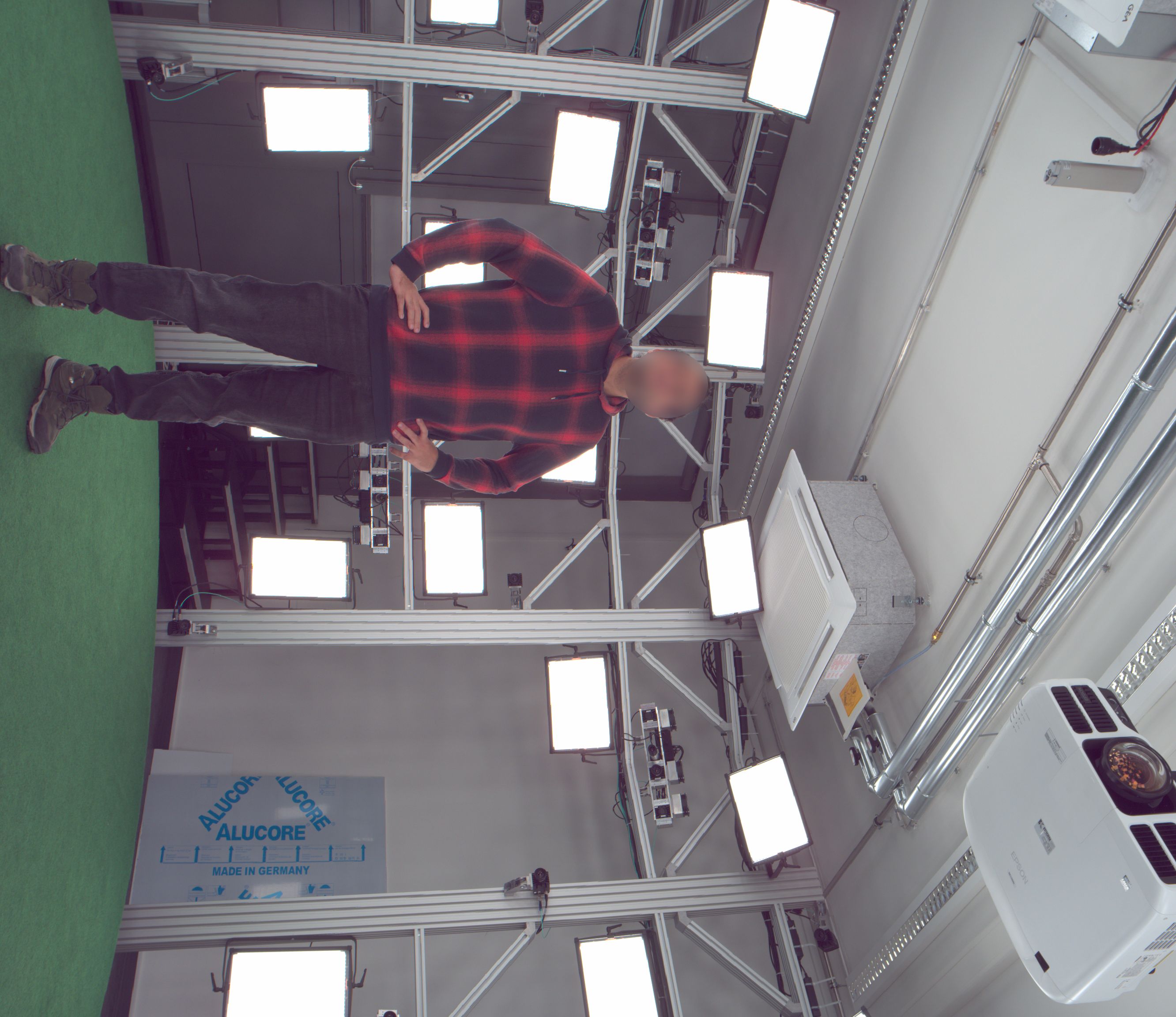}
        \includegraphics[angle=90,width=\textwidth]{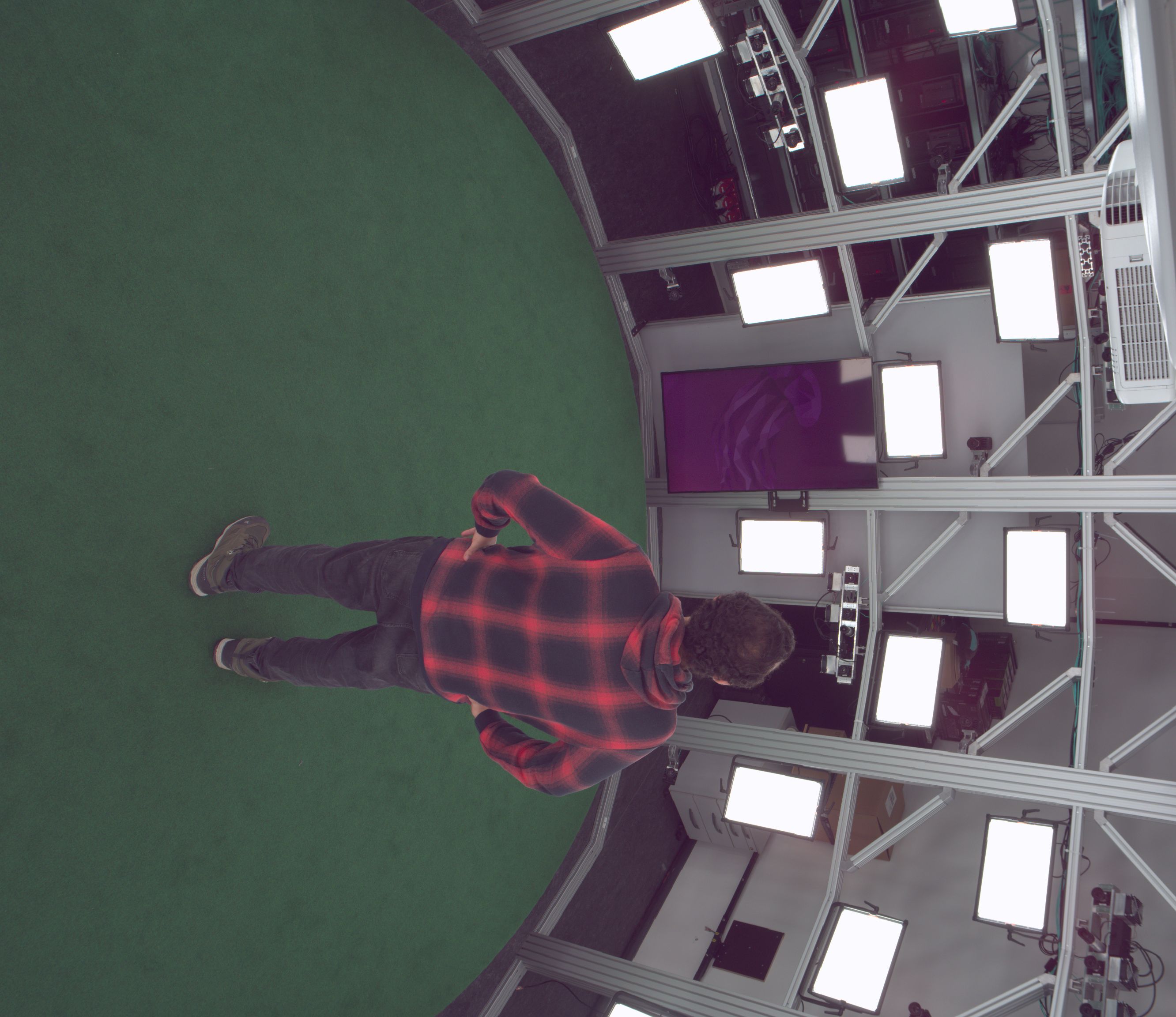}
        \caption{Input Image}
        \label{fig:subfig1}
    \end{subfigure}
    \begin{subfigure}[t]{\mywidth}
        \centering
        \includegraphics[angle=270,width=\textwidth]{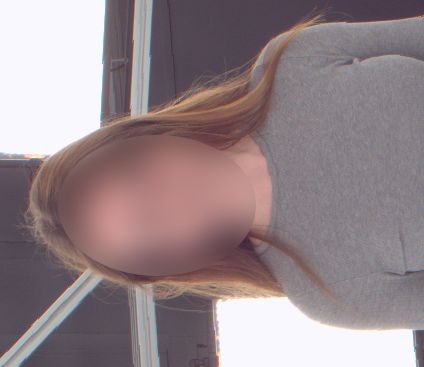}
        \includegraphics[angle=270,width=\textwidth]{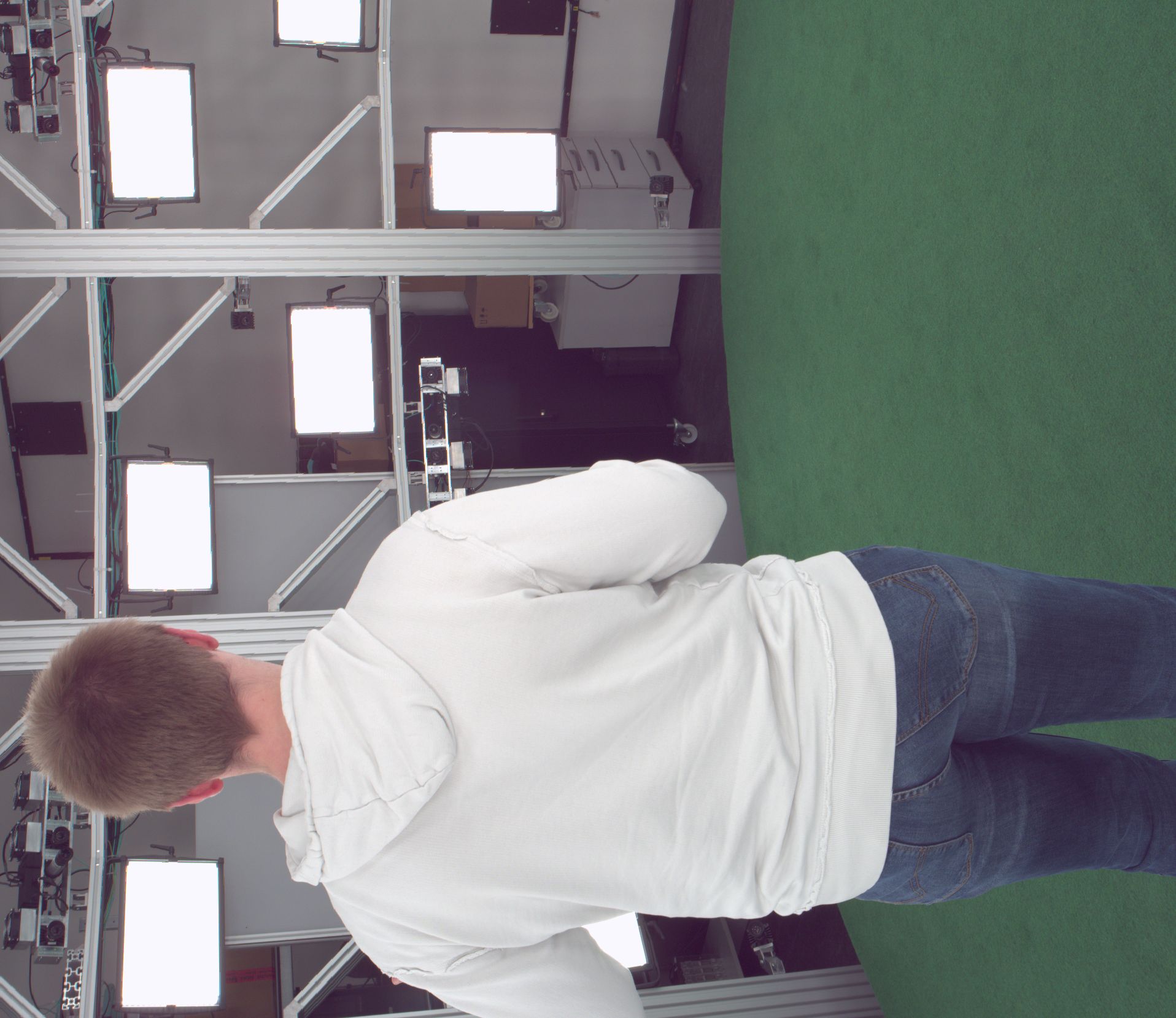}
        \includegraphics[angle=90,width=\textwidth]{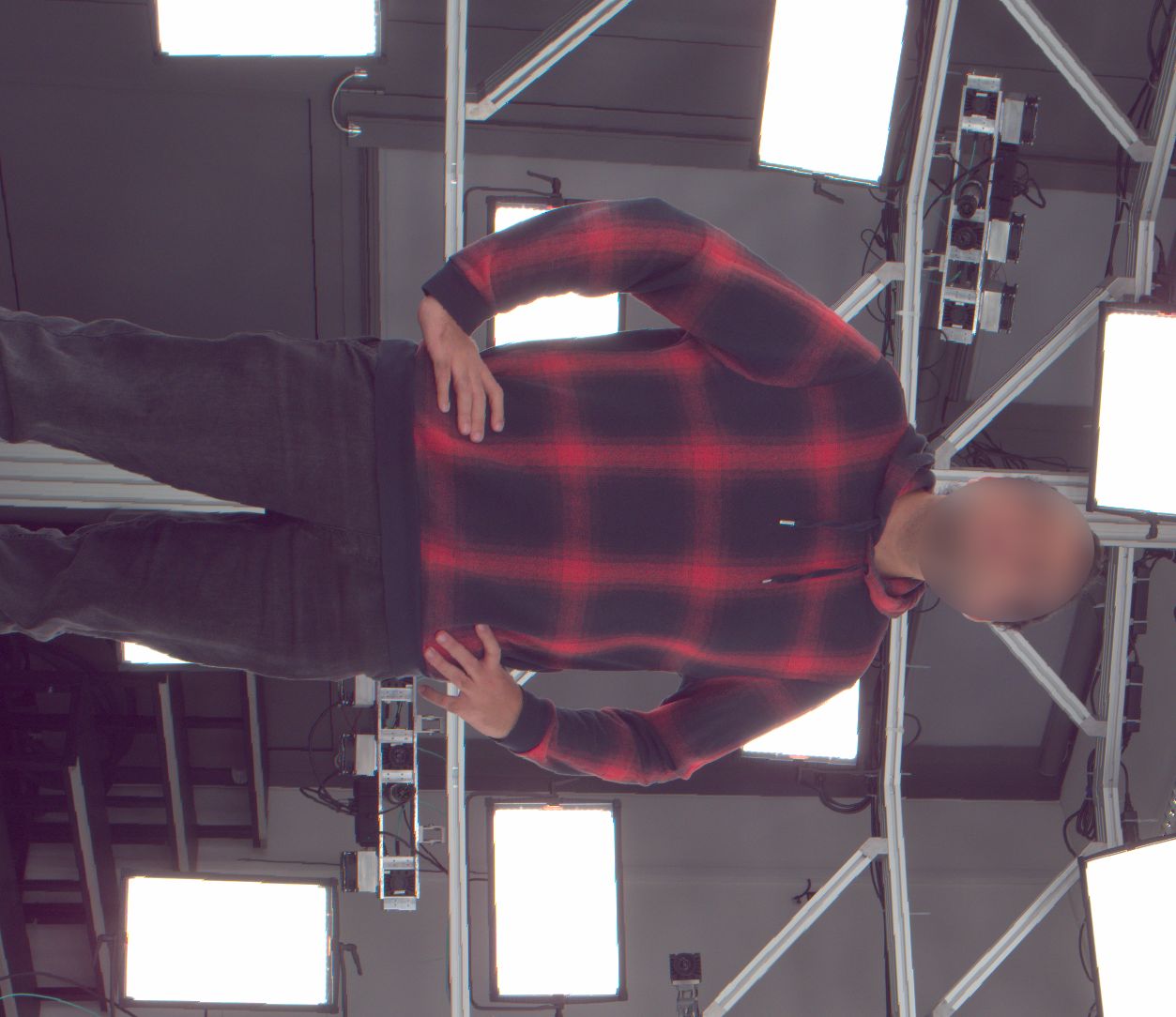} %
        \includegraphics[angle=90,width=\textwidth]{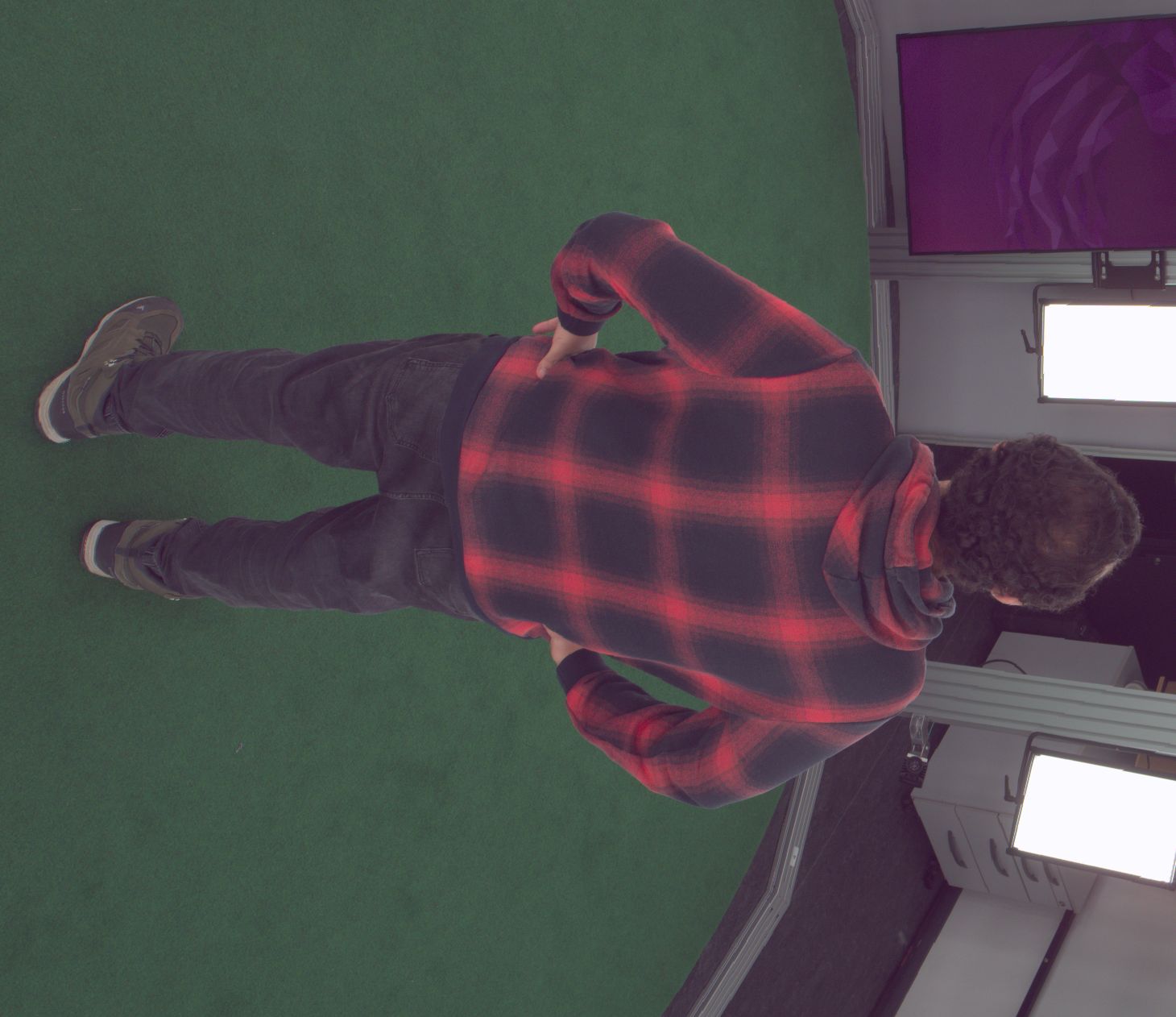}
        \caption{Zoomed-In}
        \label{fig:subfig2}
    \end{subfigure}
    \begin{subfigure}[t]{\mywidth}
        \centering
        \includegraphics[angle=270,width=\textwidth]{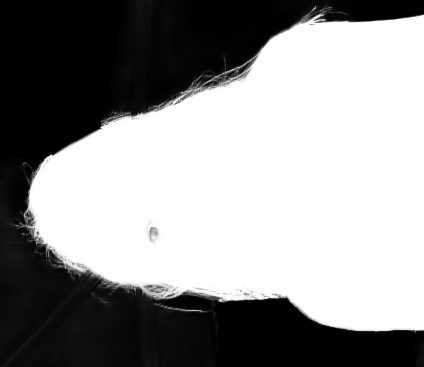}
        \includegraphics[angle=270,width=\textwidth]{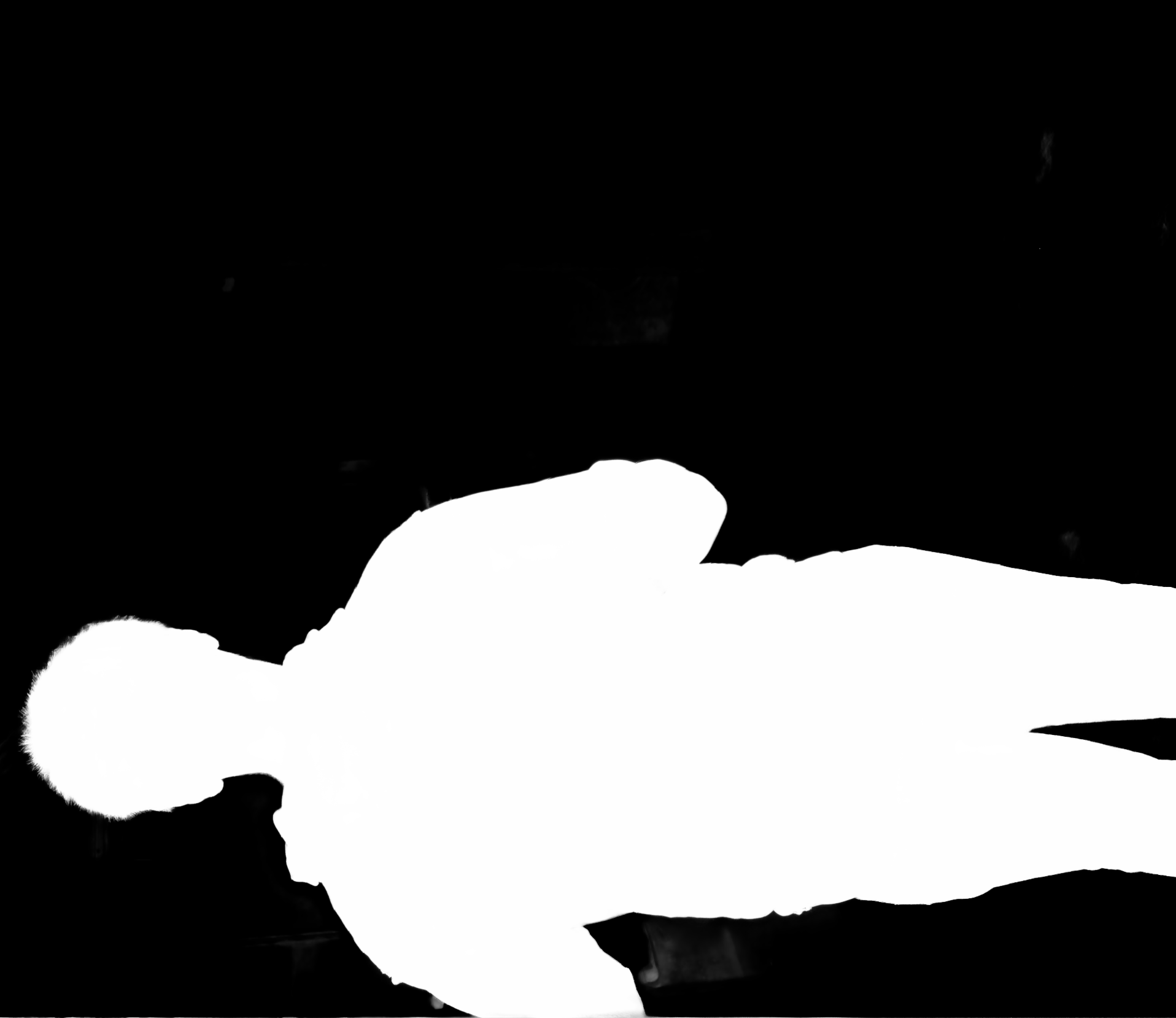}
        \includegraphics[angle=90,width=\textwidth]{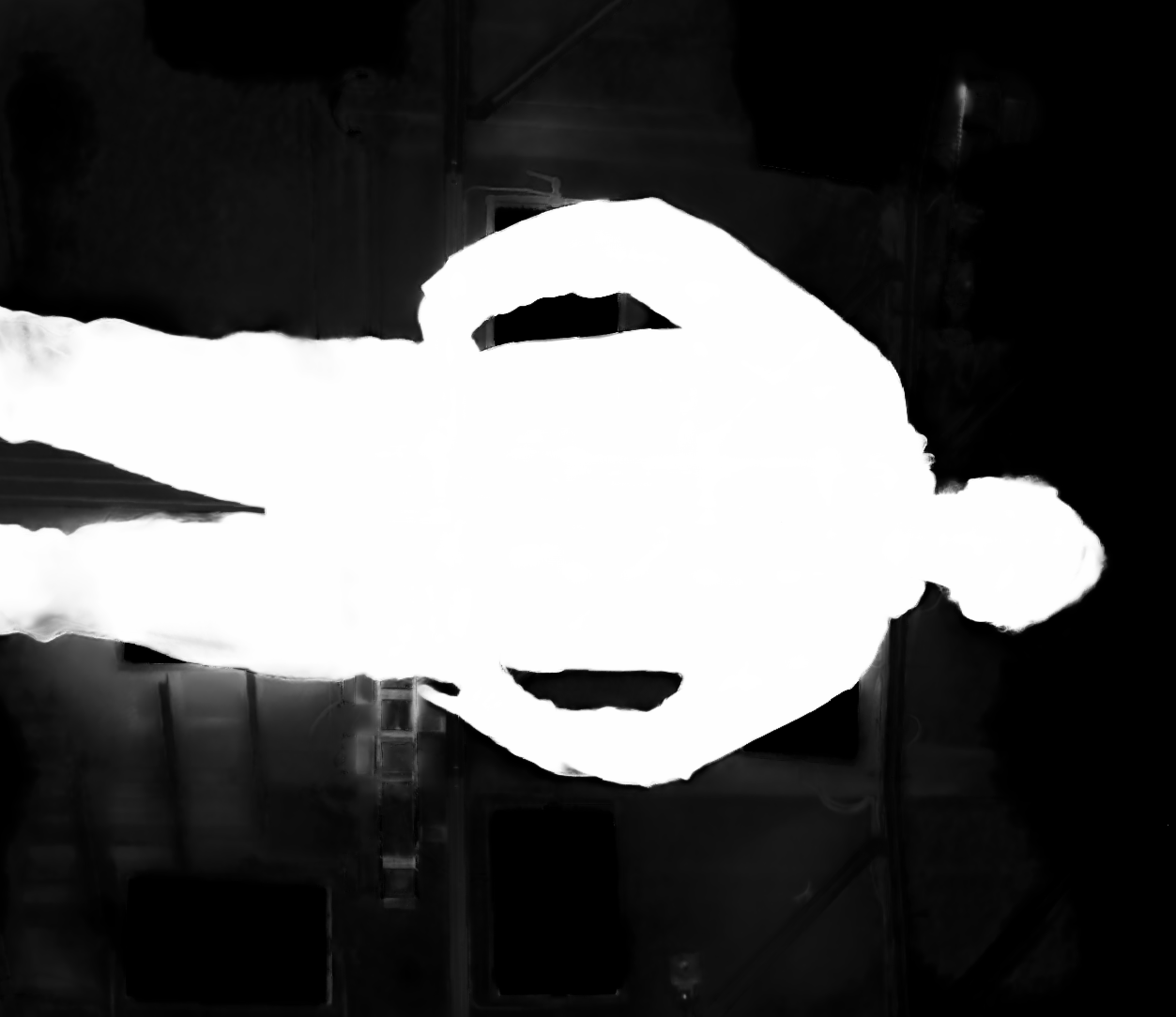}
        \includegraphics[angle=90,width=\textwidth]{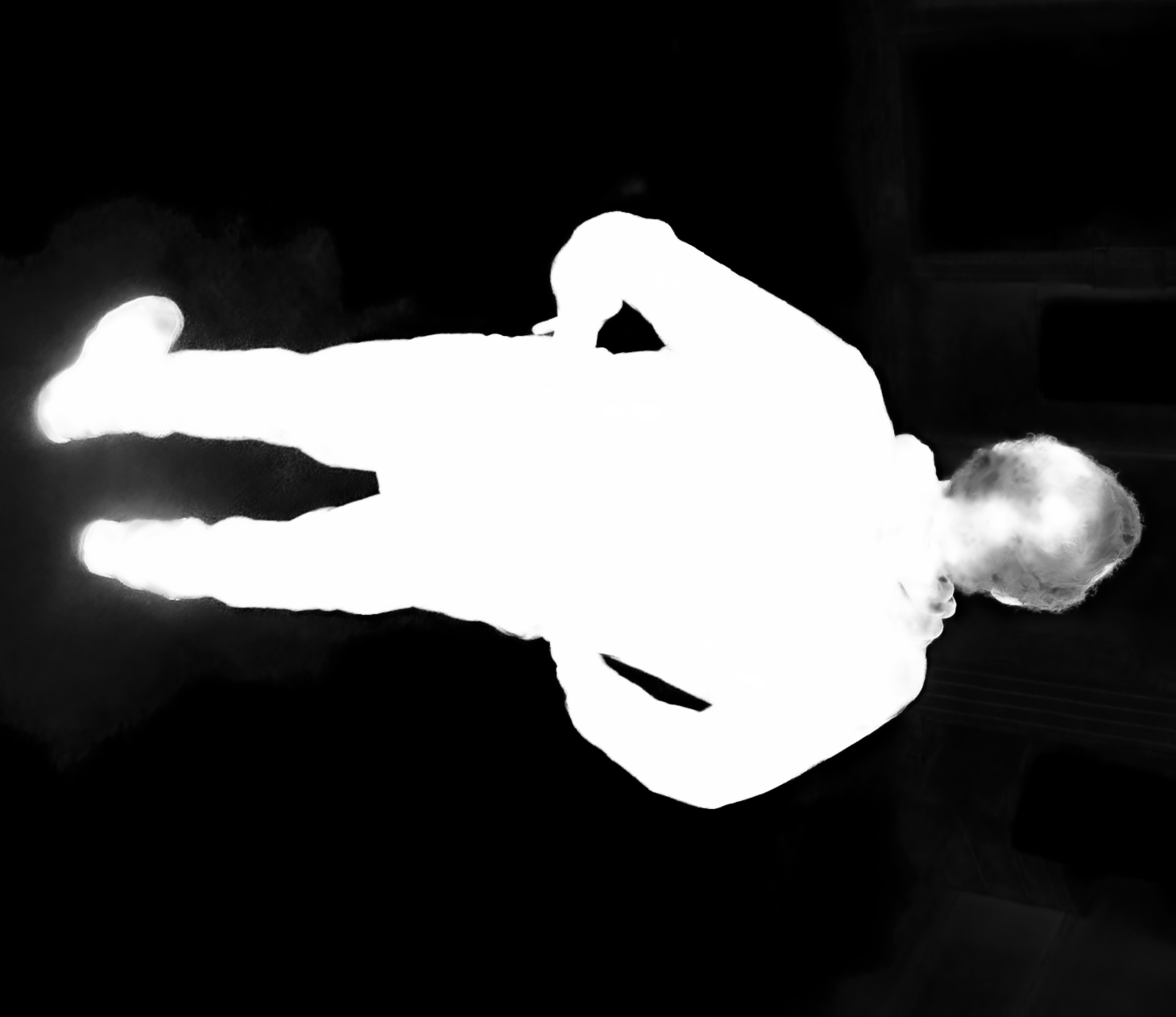}
        \caption{Teacher Base}
        \label{fig:subfig3}
    \end{subfigure}
    \begin{subfigure}[t]{\mywidth}
        \centering
        \includegraphics[angle=270,width=\textwidth]{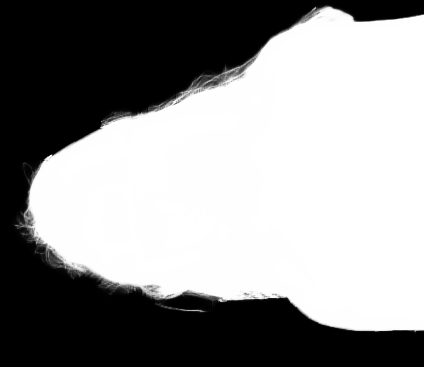}
        \includegraphics[angle=270,width=\textwidth]{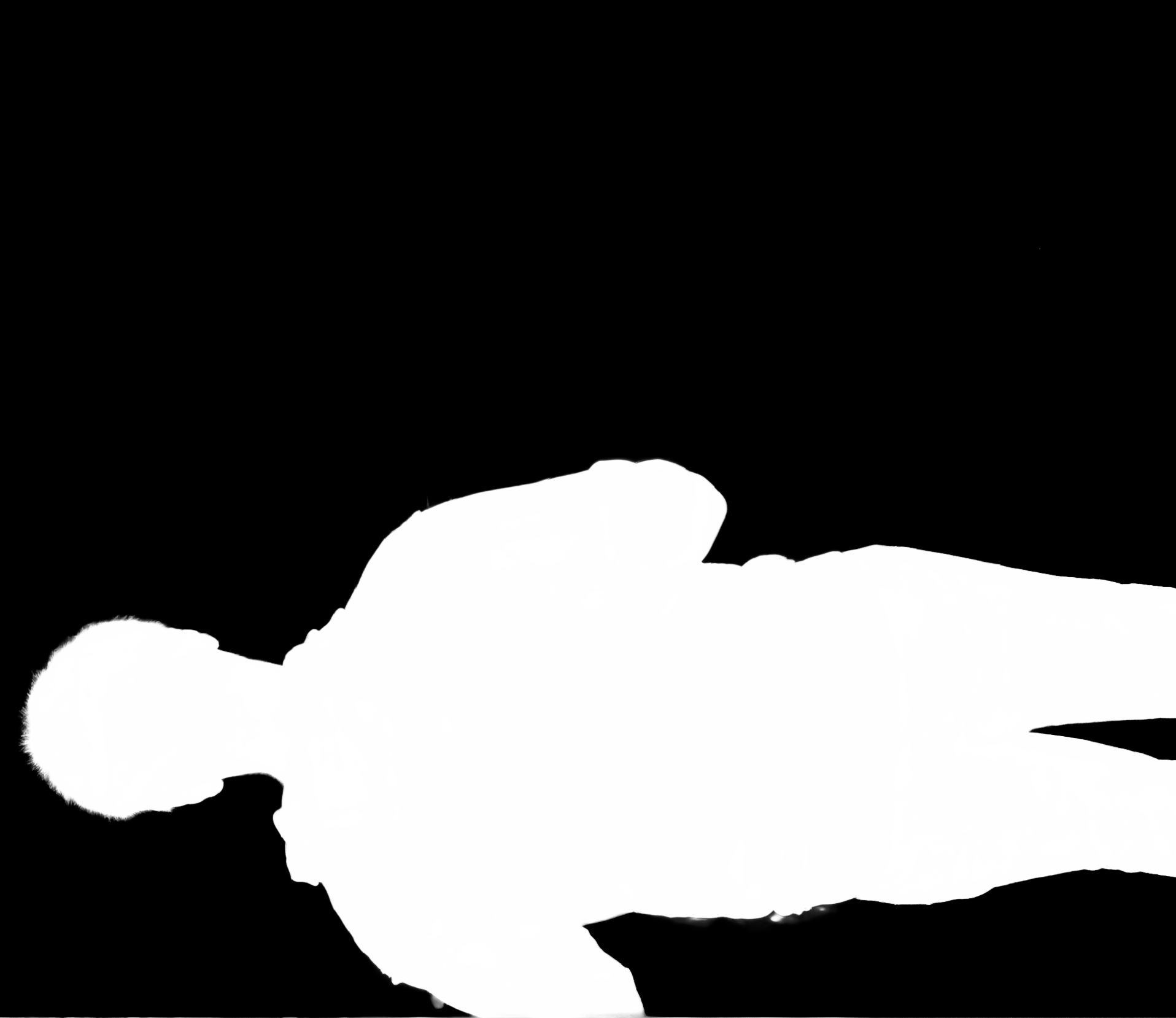}
        \includegraphics[angle=90,width=\textwidth]{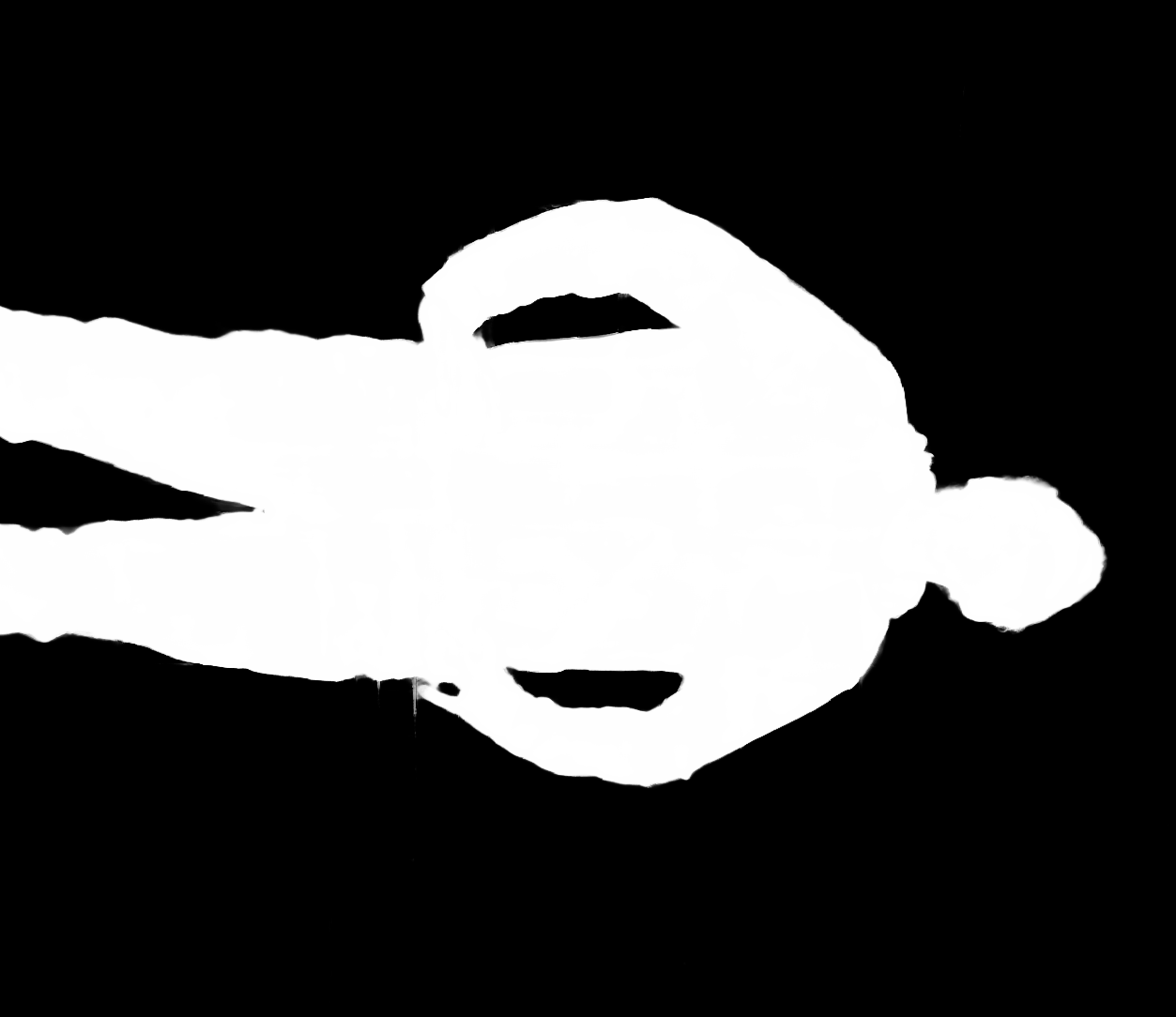}
        \includegraphics[angle=90,width=\textwidth]{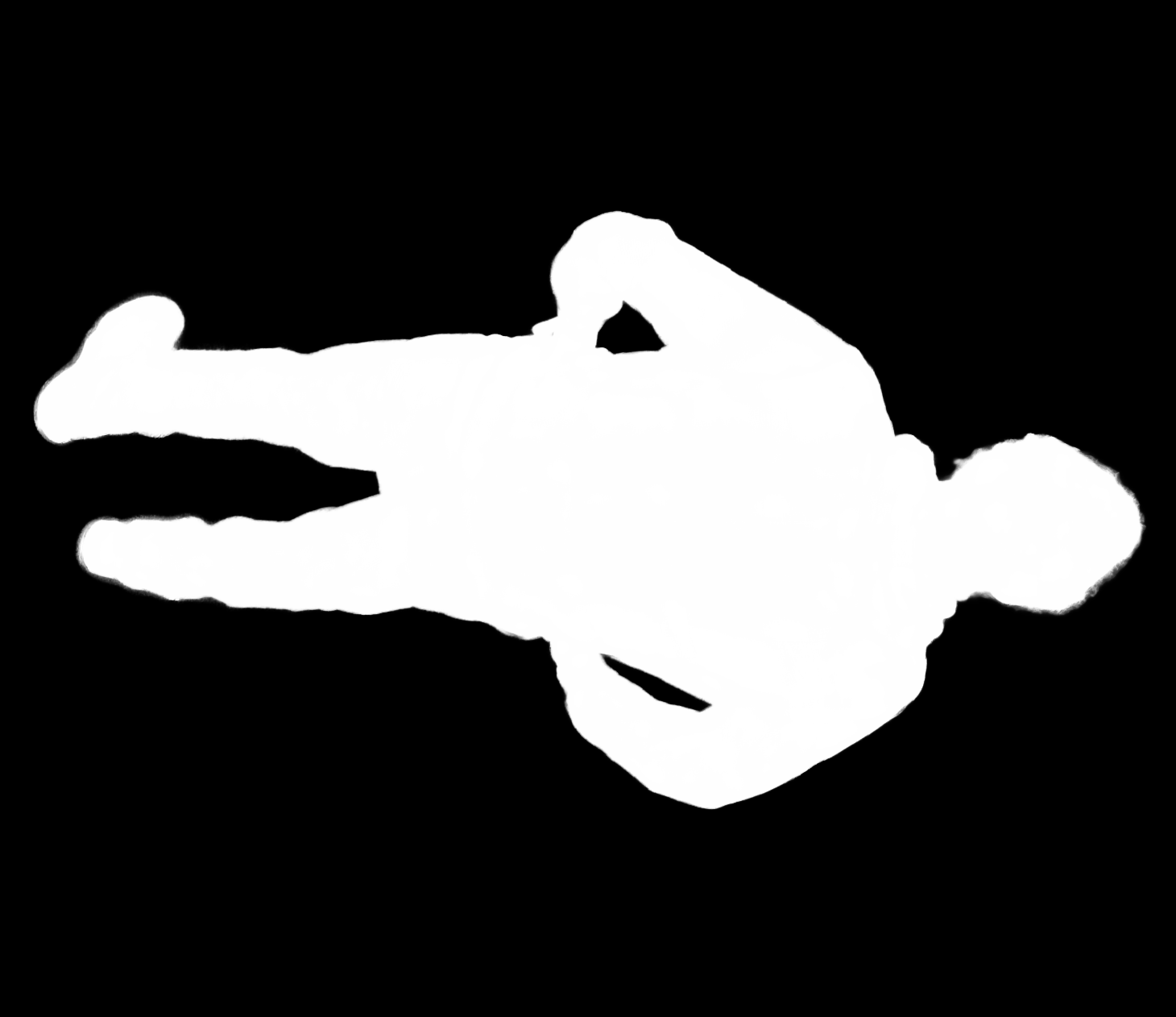}
        \caption{Teacher Refined}
        \label{fig:subfig4}
    \end{subfigure}
    \begin{subfigure}[t]{\mywidth}
        \centering
        \includegraphics[angle=270,width=\textwidth]{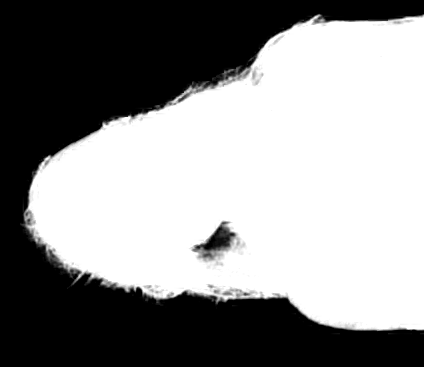}
        \includegraphics[angle=270,width=\textwidth]{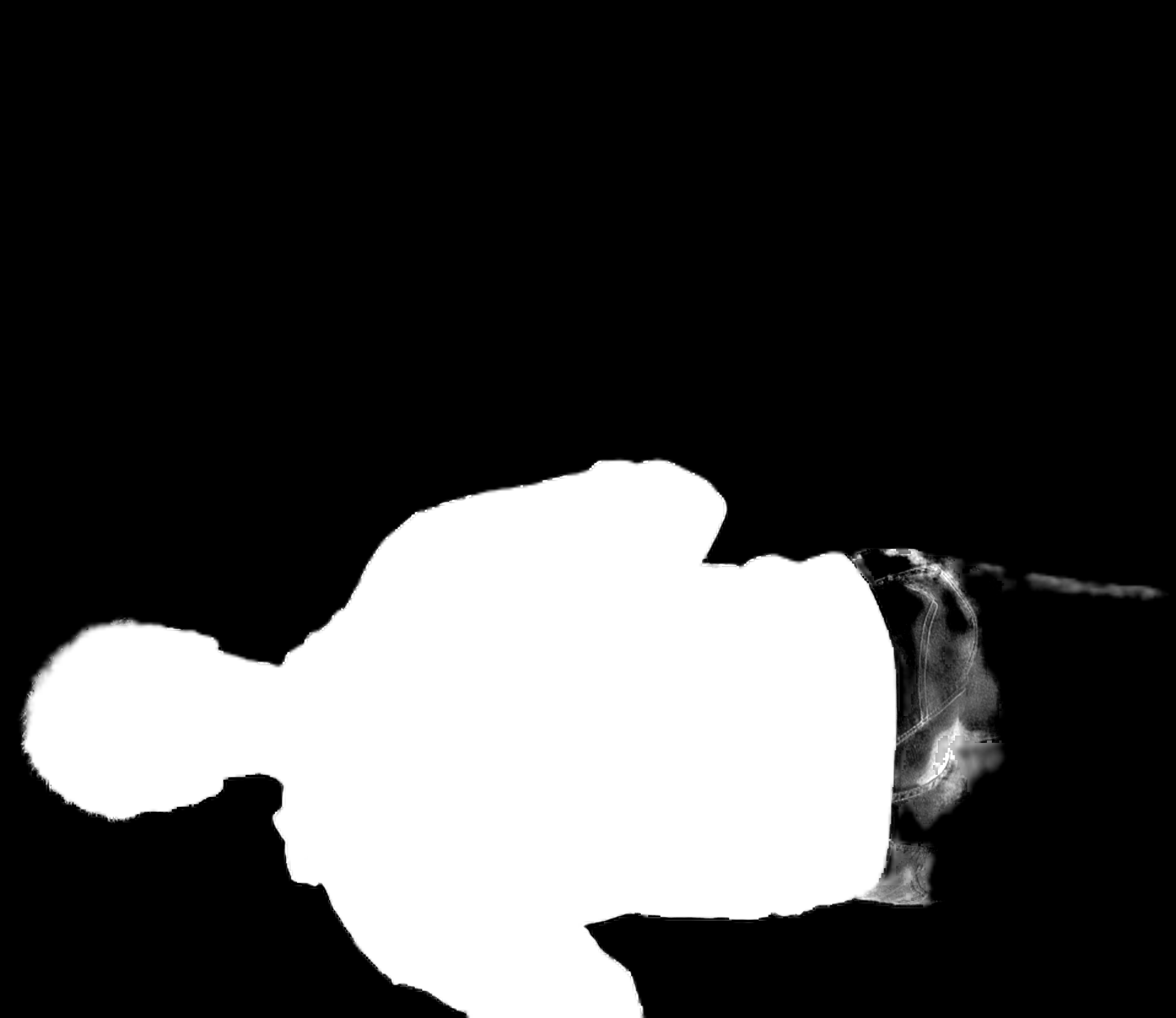}
        \includegraphics[angle=90,width=\textwidth]{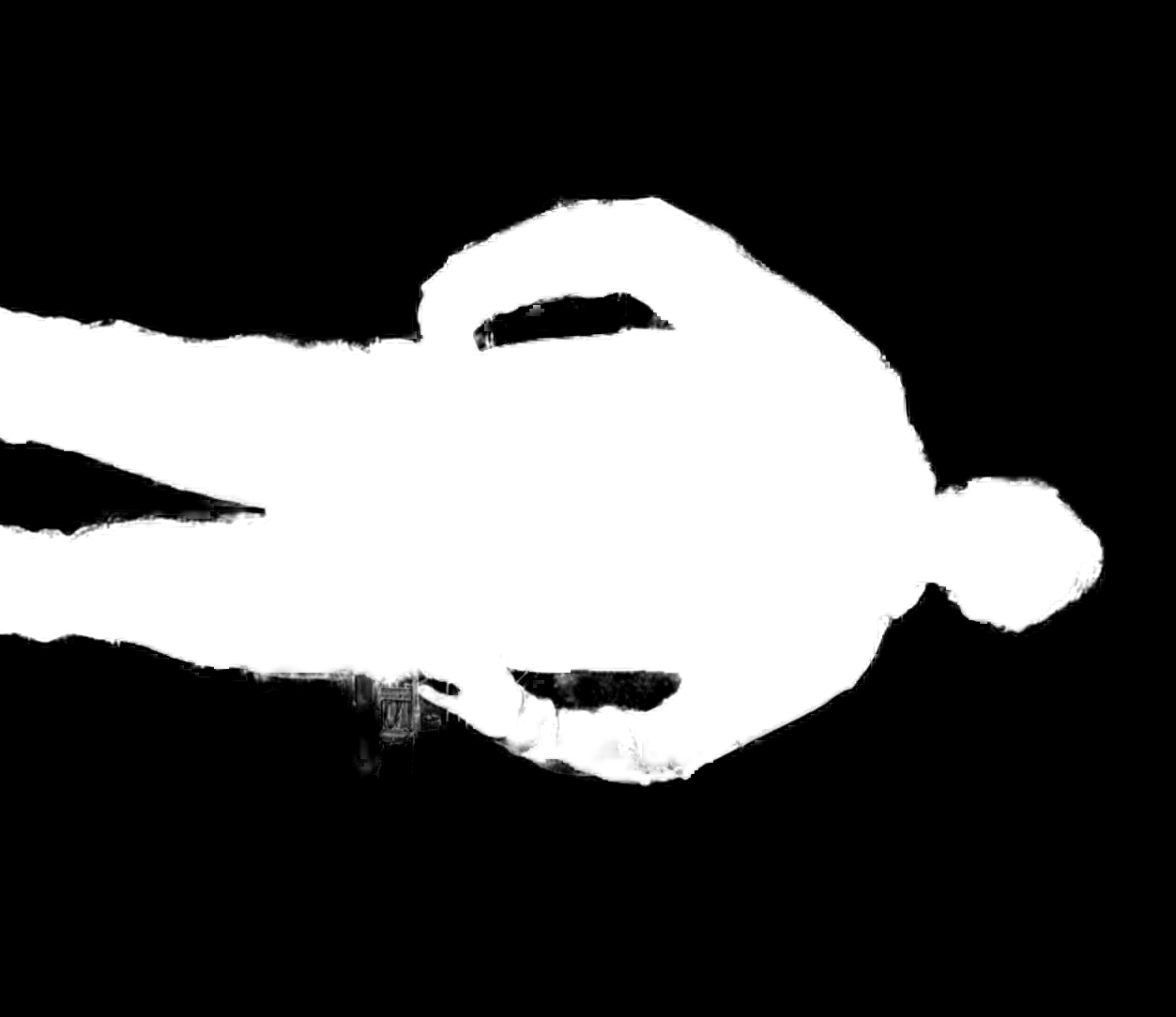}
        \includegraphics[angle=90,width=\textwidth]{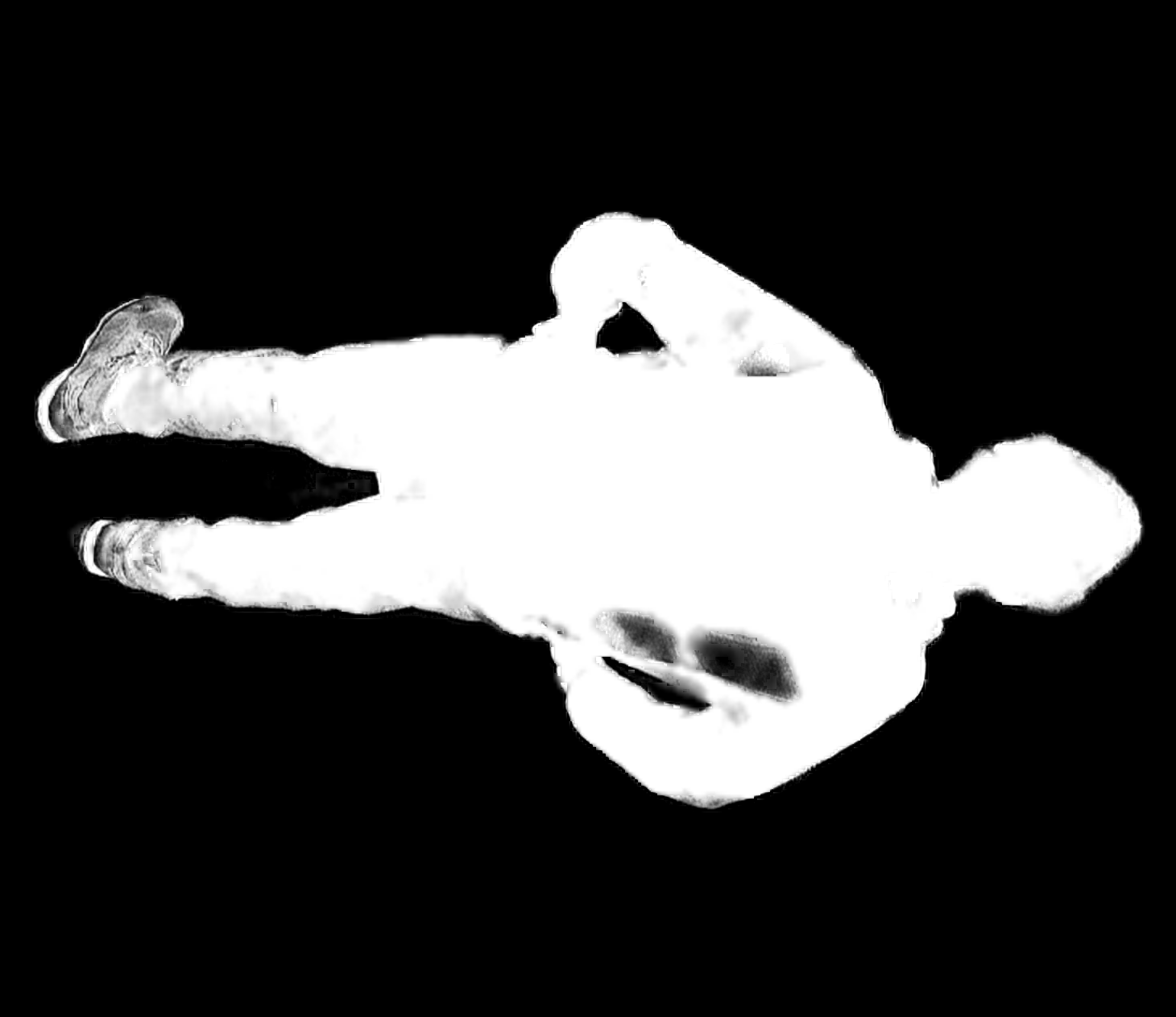}
        \caption{Student Base}
        \label{fig:subfig5}
    \end{subfigure}
    \begin{subfigure}[t]{\mywidth}
        \centering
        \includegraphics[angle=270,width=\textwidth]{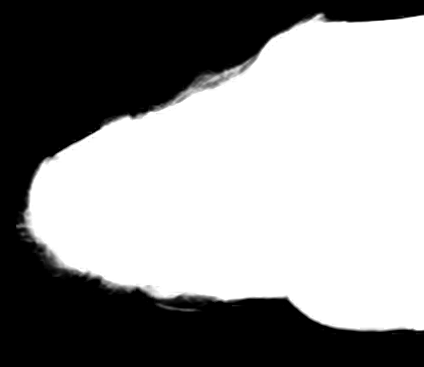}
        \includegraphics[angle=270,width=\textwidth]{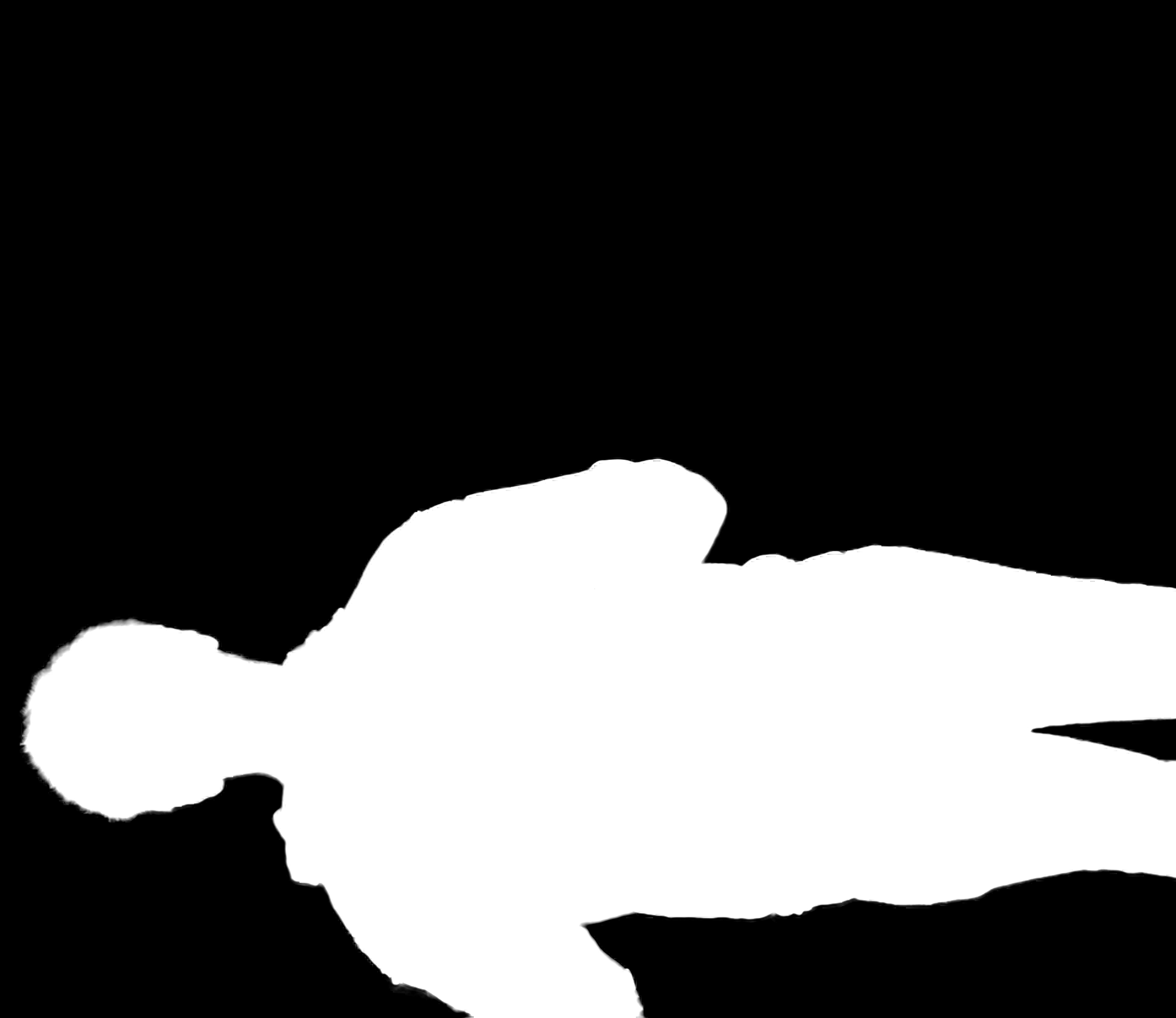}
        \includegraphics[angle=90,width=\textwidth]{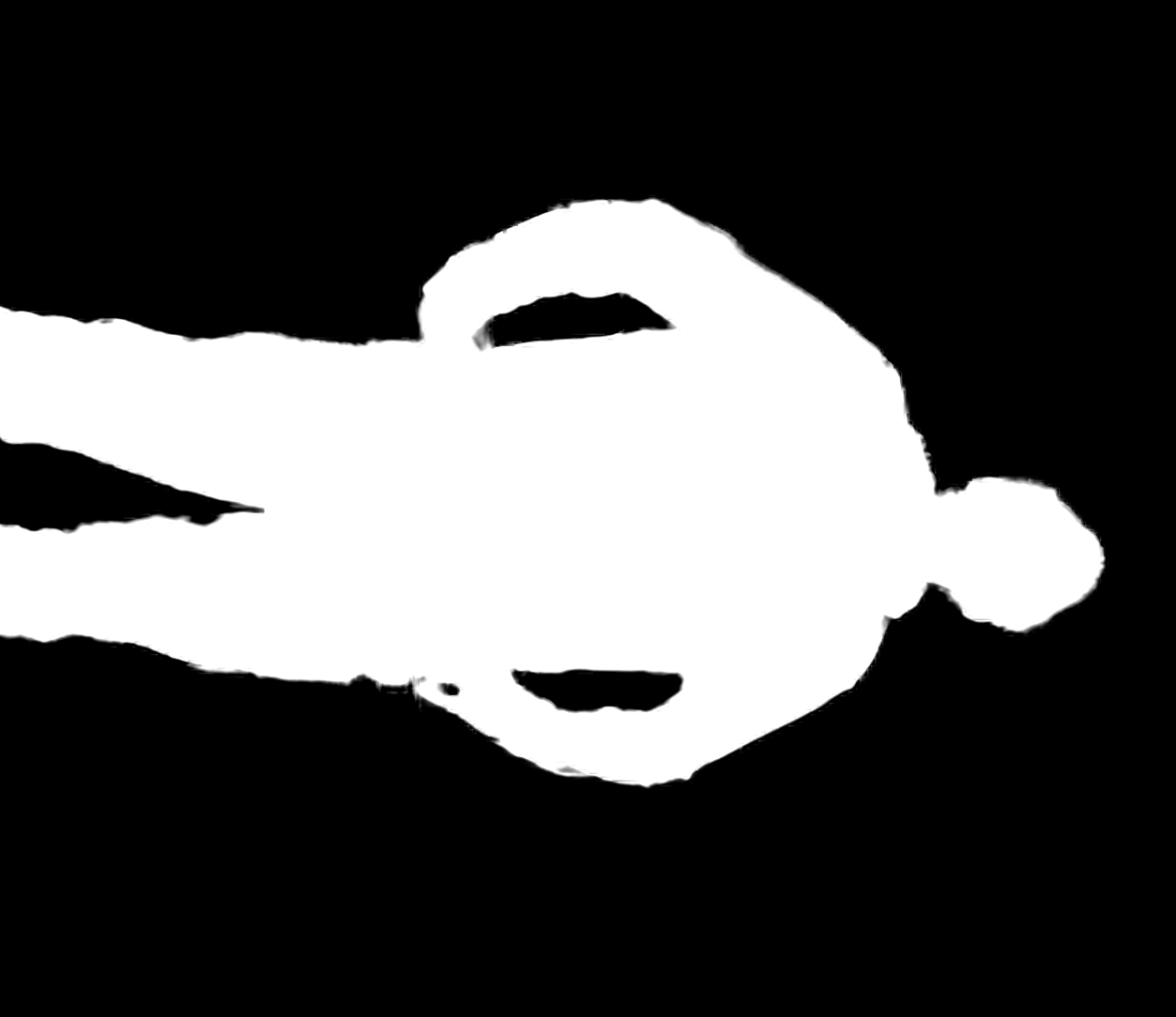}
        \includegraphics[angle=90,width=\textwidth]{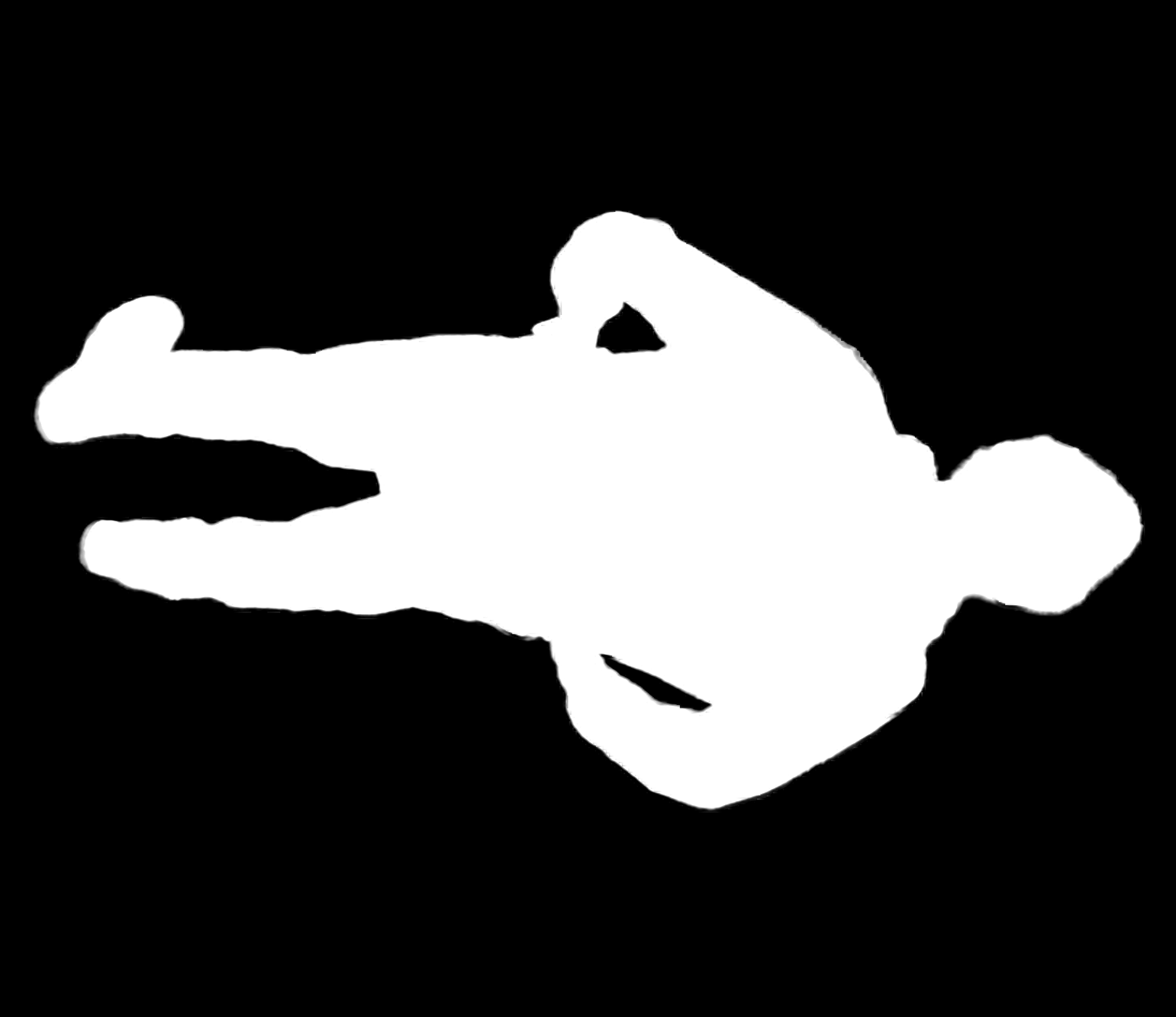}
        \caption{Student Refined}
        \label{fig:subfig6}
    \end{subfigure}
    \caption{Visual comparison of the  outputs from the teacher and student models, predicted from the input image in (a). (b) provides a zoomed-in view of a region of interest. (c) and (d) present the outputs of the teacher model {ViTMatte-S} \cite{yao2024vitmatte} in its baseline and refined configurations, respectively. Similarly, (e) and (f) show the outputs of the student model {BGMV2} \cite{lin2021real} in its baseline and fine-tuned versions.
 \vspace{-0.3cm}}
    \label{fig:qualitative}
\end{figure*}

\subsection{Baseline Methods and Comparisons}
To evaluate the performance of our proposed approach, we compare against a set of baseline methods quantitively, spanning both real-time and high-quality matting models. These include recent architectures that operate either with or without background input.

\paragraph{Quantitative Results}

We evaluate the performance of baseline matting approaches on the hold-out validation set using three standard metrics: mean squared error (MSE), sum of absolute differences (SAD), and the spatial-gradient error (Grad), all computed over full-resolution images.  The results are summarized in \cref{tab:comparison}.

First, we establish lower-bound baselines with real-time matting methods that operate without background input, relying solely on RGB data, in the top section of \cref{tab:comparison} 
These models illustrate the challenge of capture-stage matting when explicit scene context is unavailable. As expected, the absence of background cues results in consistently lower accuracy, confirming that for capture-stage applications, matting without auxiliary input is inherently limited -- particularly under real-time constraints.

The middle section presents matting methods that operate with \textit{additonal background input}. 
Several of these are adapted variants of existing models (prefixed with \textit{Bg-}) originally designed for RGB or RGB+trimap inputs but modified to accept an additional background image alongside the RGB input. 
This additional cue consistently improves performance; for example, 
augmenting DAMNet~\cite{wang2022effective} with background input reduces MSE by more than an order of magnitude (a factor of $14.6$).
Since capture-stage data differs from standard open datasets due to lighting, reflections, and shadows, %
it is essential to incorporate such data into the learning process.
In this setting, adding domain adaptation (da) to Bg-DAMNet yields only a modest MSE reduction of $18\%$.
In contrast, our scribble-based student–teacher strategy adapts naturally to capture-stage conditions by exploiting additional non-composited training images, resulting in a \textbf{substantial 61\% MSE improvement}.
This aligns with prior findings for in-the-wild matting, where improved training data was shown to significantly enhance performance \cite{li2022bridging}.

Finally, the bottom section reports results for our proposed models. Bg-ViTMatte\textsuperscript{\dag}, fine-tuned on a hybrid dataset using sparse labels, is exposed to capture-stage data during training. This model achieves the best overall MSE and SAD scores, reducing MSE by $36.2\%$ compared to its non-hybrid counterpart, highlighting the benefits of a hybrid-data training. The distilled BGMV2\textsuperscript{*} model attains the lowest Grad error while maintaining competitive accuracy at $152$ FPS on an NVIDIA GeForce RTX 4070 Ti SUPER.
This confirms that our approach scales seamlessly from offline to real-time settings, while avoiding the heavy annotation requirements of conventional training pipelines.

\begin{table}[t]
\centering
\begin{tabularx}{\linewidth}{lrrr}
\toprule
\textbf{Method} & \textbf{MSE ($\downarrow$)} & \textbf{SAD ($\downarrow$)}  & \textbf{Grad ($\downarrow$)}\\
&\footnotesize{$\cdot 10^{-4}$}&\footnotesize{$\cdot 10^{-3}$}&\footnotesize{$\cdot 10^{-5}$}\\
\midrule
MODNet\textsuperscript{*}~\cite{ke2022modnet} &$258.070$&$28.185$&$11.236$\\
VideoMatte\textsuperscript{*}~\cite{lin2021real}&$124.566$&$19.31$& $9.955$\\
BiMatting\textsuperscript{*}~\cite{qin2023bimatting}&$688.757$&$79.614$&$15.941$\\
\midrule
BGMV2\textsuperscript{*} (Base) \cite{lin2021real}     & \(20.694\) & \(3.564\) & $4.772$\\
BGM \cite{sengupta2020background} & $83.970$ & $11.806$ & $12.548$ \\
Bg-DAMNet \cite{wang2022effective} & \(63.419\) & \(18.122\) & \(23.057\)\\
Bg-DAMNet (da) \cite{wang2022effective}  \hspace{-0.3cm} & \(52.025\) & \(16.560\) & \(21.079\)\\
Bg-ViTMatte \cite{yao2024vitmatte} &  \colorbox{second}{$7.805$} & $5.760 $ & $4.221$\\
\midrule
Bg-ViTMatte\textsuperscript{\dag} (Ours) & \colorbox{best}{$4.978$} & \colorbox{best}{\(1.415\)} & \colorbox{second}{$3.978$}\\
BGMV2\textsuperscript{*} (D.) (Ours)&  \(8.054\) & \colorbox{second}{\(2.032\)} & \colorbox{best}{$3.783$}\\
\bottomrule
\end{tabularx}
\caption{
Matting accuracy on capture-stage data. The best results are highlighted in red, the second best in orange. \textsuperscript{*} denotes real-time methods, and \textsuperscript{\dag} indicates models trained on a hybrid dataset. BGMV2\textsuperscript{*} (D.) is the  student model trained via knowledge distillation. The top section lists methods without background information, the middle section includes those using background information, and the bottom section shows our proposed methods.
}
\label{tab:comparison}
\end{table}

\paragraph{Qualitative Evaluation}
To further analyze the impact of the task-specific refinements, we performed a qualitative comparison of both the teacher and student models before and after these refinements. %
From \cref{fig:qualitative} we can observe that the teacher model tends to produce false positives in its base configuration, which is mitigated through the refinement. Meanwhile, the student model rather struggles to capture all foreground regions accurately in its base setup. Fine-tuning on the teacher-generated data, however, significantly improves its performance by more precisely outlining foreground elements and reducing errors.

\subsection{Ablation Study on Student Model Training}\label{sec:student_train_direct}

To assess the effectiveness of different training strategies for real-time matting under limited supervision, we conduct an ablation study comparing several training paradigms for the student model (BGMV2). We evaluate how each variant performs in terms of alpha mask accuracy and consistency with the teacher model.
\begin{table}[h]
\centering
\caption{Ablation of student model training strategies. Performance is measured on the capture stage validation set.}
\begin{tabular}{lcc}
\toprule
\textbf{Training Strategy} & \textbf{MSE} ($\downarrow$) & \textbf{SAD} ($\downarrow$) \\
\midrule
Scribble supervision only        & 89.2 $\cdot 10^{-4}$ & 10.4 $\cdot 10^{-3}$ \\
Frozen refinement (2-stage)      & 27.8 $\cdot 10^{-4}$ & 5.75 $\cdot 10^{-3}$ \\
{Knowledge distillation} & {8.05 $\cdot 10^{-4}$} & {2.03 $\cdot 10^{-3}$} \\
\bottomrule
\end{tabular}
\label{tab:student_training_ablation}
\end{table}
\paragraph{Training Strategy Comparison}
We first fine-tune the base model of BGMV2 \cite{lin2021real} directly on our hybrid dataset combining Adobe DIM with the 41
scribble-annotated failure cases. We observe poor performance, especially when the refinement stage is also trained on sparse labels, as shown in \cref{tab:student_training_ablation}. %
To mitigate the shortcomings of sparse labels, we freeze the refinement module and train only the base network on sparse data, then fine-tune the refinement network on the Adobe dataset. While this improves performance, it still underperforms compared to teacher-supervised training.
Finally, we fine-tune the student model using full-resolution alpha masks generated by the refined teacher model on 580 new capture stage images. This method leverages strong pseudo-ground-truth and results in significantly improved metrics. %

\begin{figure*}[t]
    \centering
    \begin{subfigure}[b]{0.32\linewidth}
        \centering
        \includegraphics[width=0.49\linewidth, trim=0 0 0 0, clip]{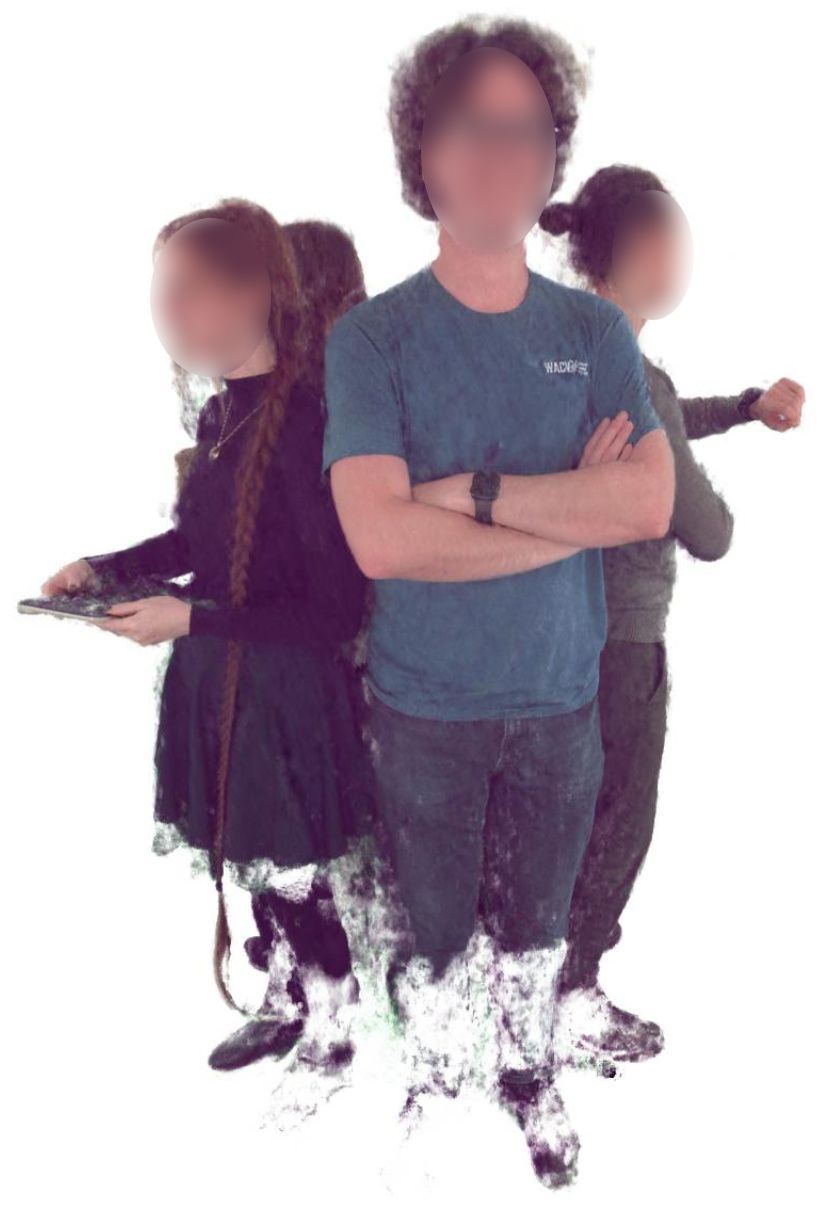}
        \includegraphics[width=0.49\linewidth]{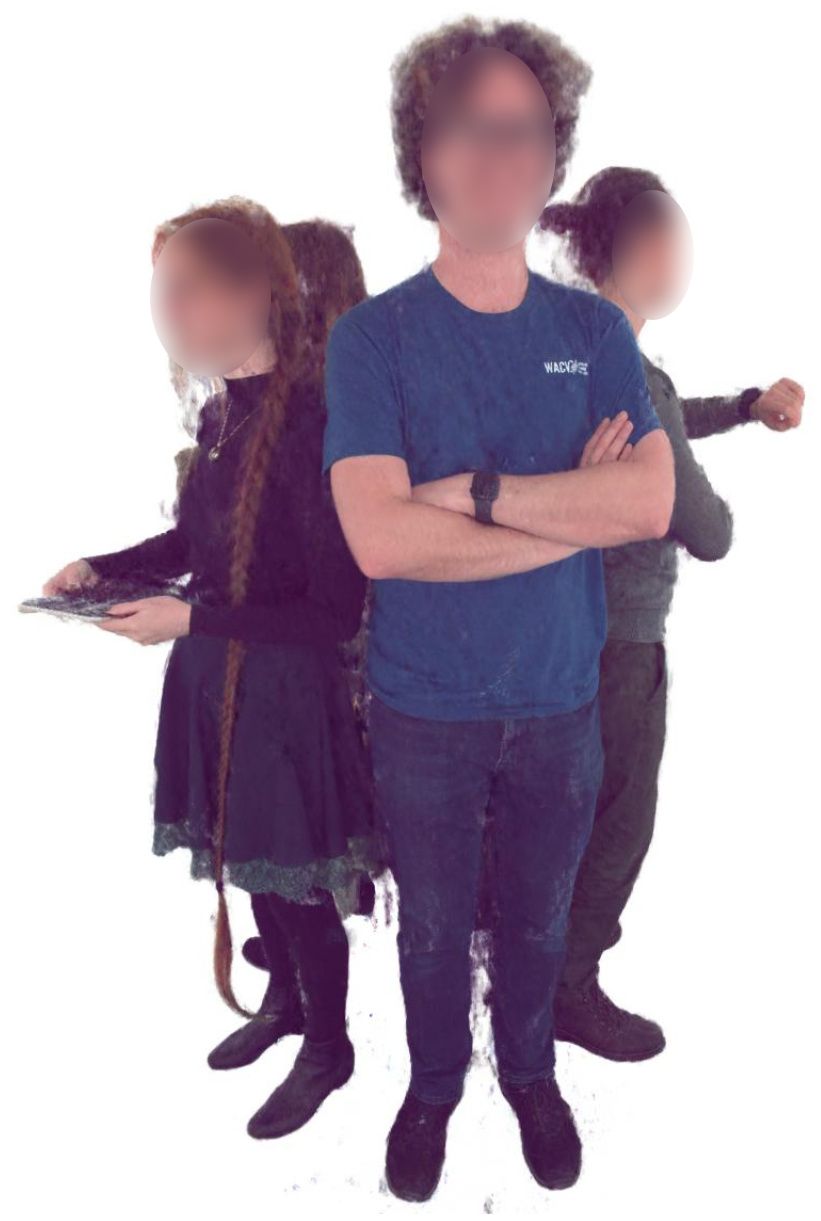}
    \end{subfigure}
    \hfill
    \begin{subfigure}[b]{0.32\linewidth}
        \centering
        \includegraphics[width=0.49\linewidth, trim=0 0 0 0, clip]{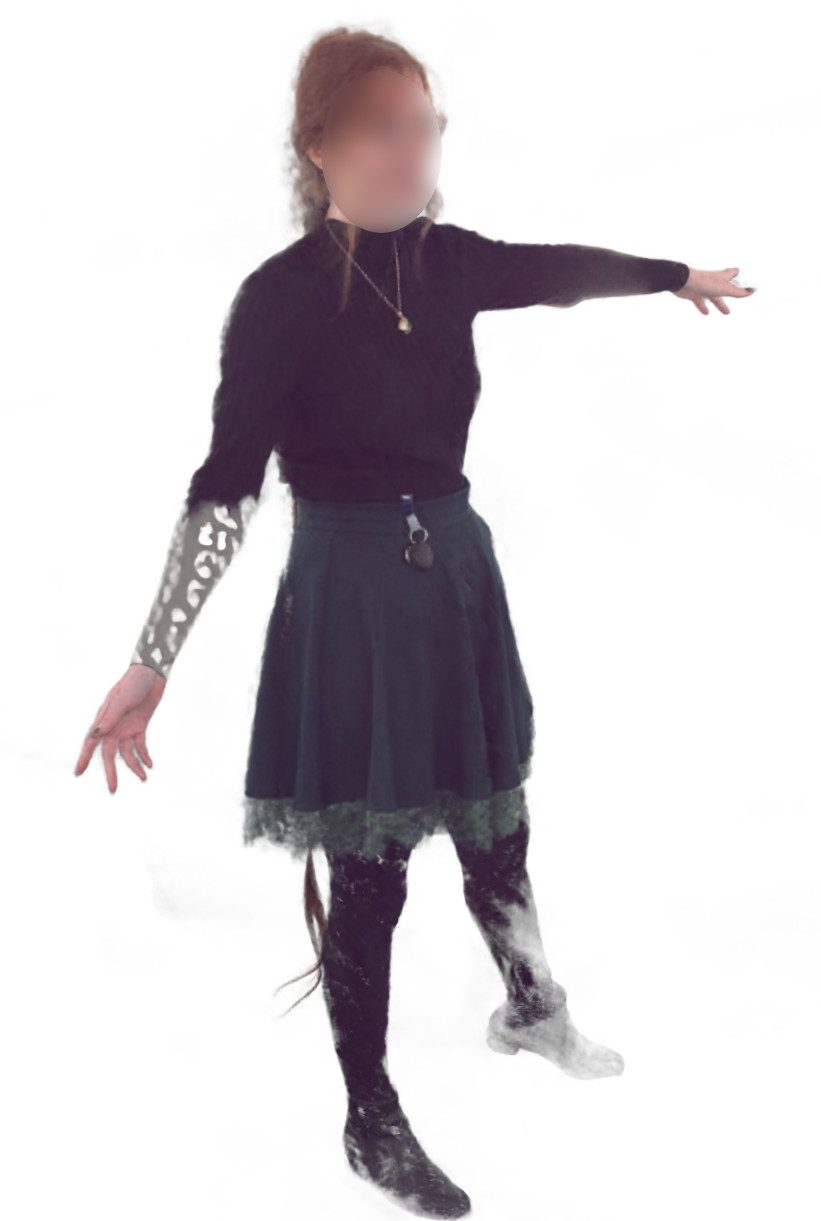}
        \includegraphics[width=0.49\linewidth]{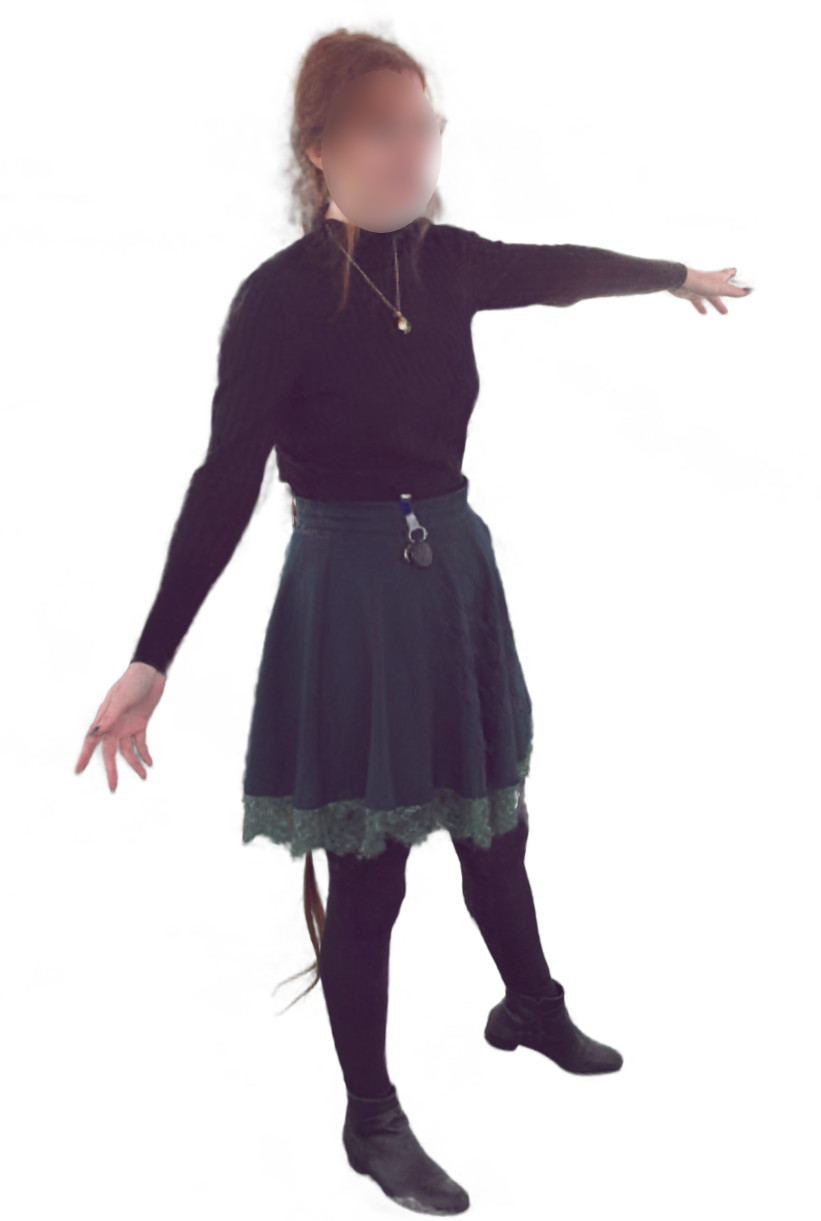}
    \end{subfigure}
    \hfill
    \begin{subfigure}[b]{0.32\linewidth}
        \centering
        \includegraphics[width=0.49\linewidth, trim=0 0 0 0, clip]{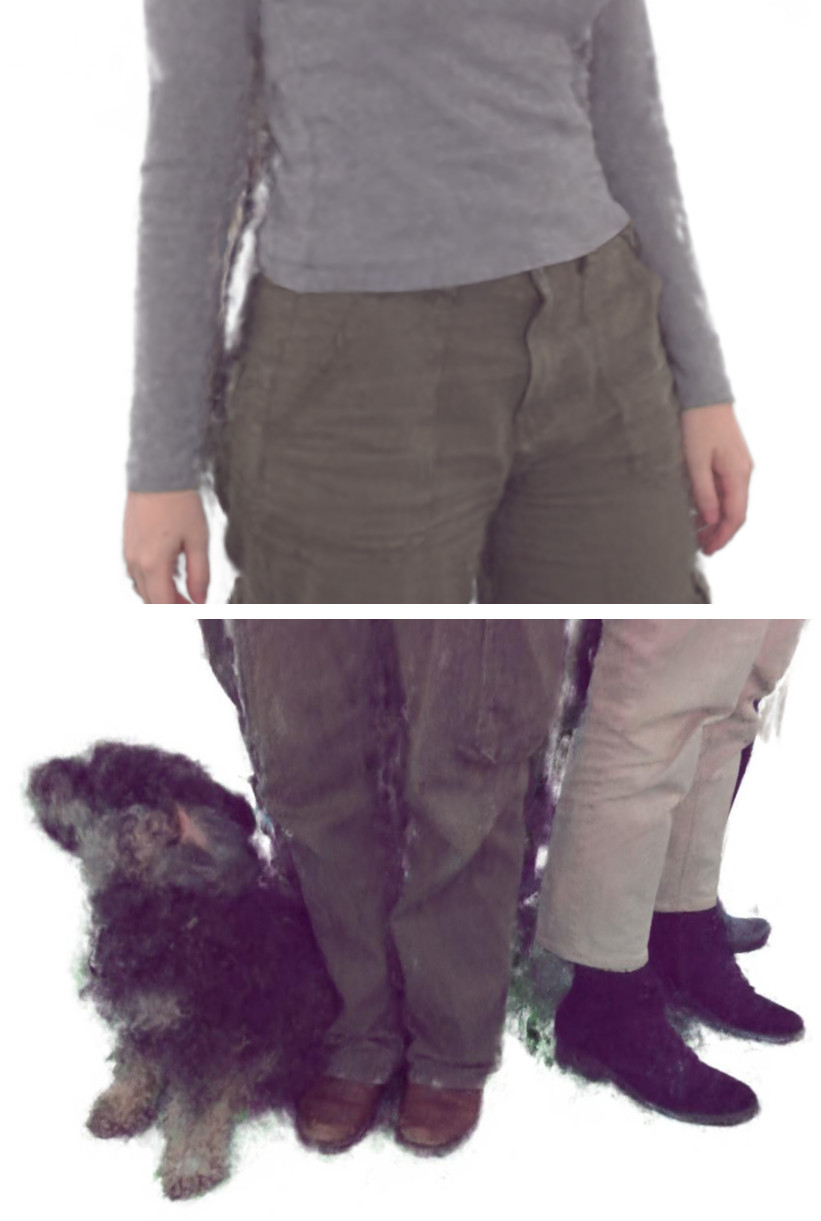}
        \includegraphics[width=0.49\linewidth]{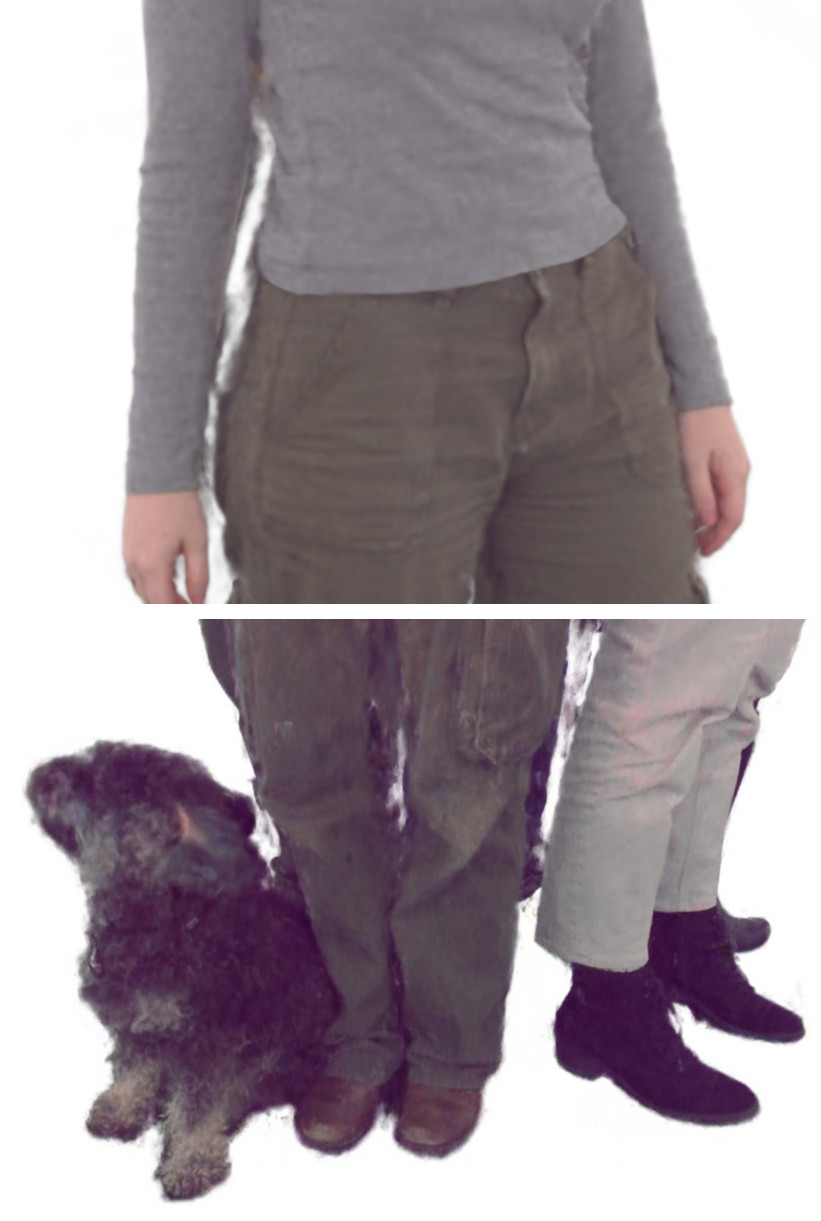}
    \end{subfigure}
    \caption{
    Qualitative comparisons of novel views from InstantNGP using the original alpha masks from BGMV2~\cite{lin2021real} (left) and our improved masks (right). 
      The baseline's erroneous matting cuts out part of the content and produces unfaithful colors to compensate for the wrong inputs.
      The refined matting yields a notable quantitative improvement, with PSNR increasing from 30.5 dB to 33.5 dB.
    }
    \label{fig:instantngp}
\end{figure*}

\paragraph{Teacher-Student Alignment Analysis}
To assess the learning capacity of the student model, we additionally compare its alpha mask predictions directly against those produced by the refined teacher model on a held-out validation set of 519 capture stage images. This evaluation focuses on alignment quality, rather than absolute accuracy against ground truth, and helps confirm whether the student successfully learns structural cues from the teacher.
Table~\ref{tab:teacher_student_alignment} shows a consistent improvement in similarity between student and teacher outputs after distillation.
These results indicate that the student effectively approximates the teacher's outputs, even under real-world capture conditions, validating the success of our distillation-based training pipeline.

\begin{table}[t]
\centering
\begin{tabular}{lrrr}
\toprule
\textbf{} & \textbf{MSE ($\downarrow$)} & \textbf{SAD ($\downarrow$)} & \textbf{Grad ($\downarrow$)} \\  %
\textbf{} &\footnotesize{$\cdot 10^{-4}$}   & \footnotesize{$\cdot 10^{-3}$} & \footnotesize{$\cdot 10^{-5}$} \\  %
\midrule
BGMV2 (Base)  &17.892 &  3.209& 6.068 \\  %
BGMV2 (Distill.)  & 9.098 & 2.126 & 5.126 \\  %
\bottomrule
\end{tabular}
\caption{Quantitative comparison of the student model's  %
performance against the teacher model %
on validation data. The results highlight the alignment between the student and teacher models in terms of predicted mask quality.}
\label{tab:teacher_student_alignment}
\end{table}

\subsection{Downstream Application}

To demonstrate the impact of our proposed matting pipeline on a downstream application, we integrate the predictions of the real-time matting model into a neural radiance field (NeRF) framework \cite{mildenhall2020nerf}, which generates detailed 3D scene reconstructions from posed images. Here, correct alpha masks used in pre-processing to separate the foreground are essential to achieve a precise reconstruction. %

We demonstrate the improvements of our fine-tuned matting approach on a reconstruction of our capture stage content by optimizing an \textit{Instant-NGP} model \cite{mueller2022instant} provided by Nerfstudio \cite{nerfstudio} in its default setting for the bounded model with a maximum grid resolution of $2^{16}$, and optimized for $100\,000$ iterations.
\cref{fig:instantngp} shows the resulting reconstruction, using the base version of BGMV2 \cite{lin2021real} and its fine-tuned counterpart, demonstrating that the fine-tuned model achieves a more faithful reconstruction of the scene. %

\begin{figure}[t]
    \centering
    \includegraphics[width=0.23\textwidth]{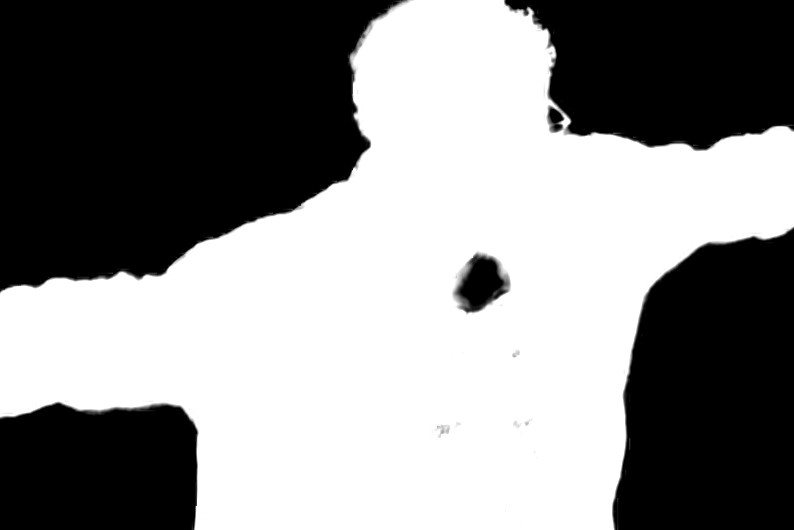}
    \includegraphics[width=0.23\textwidth]{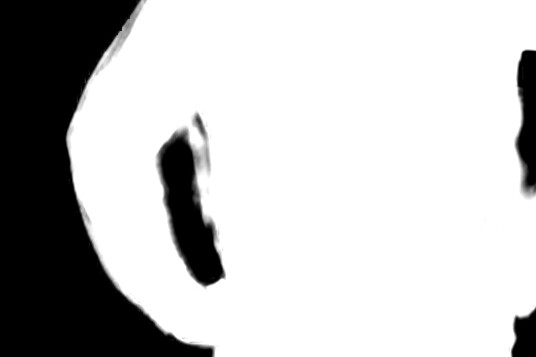}
    \caption{Potential failure cases of the matting prediction that can occur despite fine-tuning.}
    \label{fig:failure_cases}
\end{figure}
\section{Limitations}
While our method  has shown improvements in matting quality, it does not always guarantee accurate alpha mask predictions due to the inherent limitations of learning-based approaches. See \cref{fig:failure_cases} for example failure cases of the fine-tuned {BGMV2} model, showing mispredictions at object borders and an incorrect assignment of a foreground region appearing as a  ``hole'' in the foreground.

\section{Conclusion}

In this work, we addressed the unique challenges faced by data-driven matting methods in capture-stage environments, and proposed
guidelines for designing systems that better support accurate matting.
Alongside technical considerations, we introduced a lightweight knowledge-distillation pipeline that uses sparse scribble annotations to enhance a high-capacity offline model and transfer its accuracy to a real-time implementation, blending offline precision  and real-time efficiency. 
Through quantitative evaluation and integration into a NeRF reconstruction pipeline, we demonstrated both significant improvements in alpha-mask estimation and downstream applications.

\section*{Acknowledgments}

This work has been funded by the Ministry of Culture and Science North Rhine-Westphalia under grant number PB22-063A (InVirtuo 4.0: Experimental Research in Virtual Environments), and by the state of North RhineWestphalia as part of the Excellency Start-up Center.NRW (U-BO-GROW) under grant number 03ESCNW18B, and additionally by the Federal Ministry of Education and Research (BMBF) under grant no. 01IS22094A WEST-AI.

{
    \small
    \bibliographystyle{ieeenat_fullname}
    \bibliography{chapters/egbib}
}

\end{document}